\documentclass[10pt,a4paper,dvipdfmx]{article}
\usepackage{jcappub}
\usepackage{color}
\usepackage{amsmath,amssymb}

\newcommand{\LL}{ {\cal L}}
\newcommand{\kk}{ {\vec k}}
\newcommand{\pp}{ {\vec p}}
\newcommand{\rr}{ {\vec r}}

\newcommand{\xx}{ {\vec x}}
\newcommand{\sr}{ {\vec s}}
\newcommand{\vv}{ {\vec v}}

\begin{document}

\title{Will Kinematic Sunyaev-Zel'dovich Measurements Enhance the Science Return
from Galaxy Redshift Surveys?}

\author[a,b,c]{Naonori S. Sugiyama,}
\author[a,d]{Teppei Okumura,}
\author[e]{and David N. Spergel}

\affiliation[a]{Kavli Institute for the Physics and Mathematics of the Universe (WPI),
		    Todai Institutes for Advanced Study, The University of Tokyo, Chiba 277-8582, Japan}
\affiliation[b]{CREST, Japan Science and Technology Agency, Kawaguchi, Saitama, Japan}
\affiliation[c]{Department of Physics, School of Science, The University of Tokyo, Tokyo 113-0033, Japan}
\affiliation[d]{Institute of Astronomy and Astrophysics, Academia Sinica, P. O. Box 23-141, Taipei 10617, Taiwan}
\affiliation[e]{Department of Astrophysical Sciences, Princeton University, Peyton Hall, Princeton NJ 08544-0010, USA}

\emailAdd{nao.s.sugiyama@gmail.com}
\emailAdd{tokumura@asiaa.sinica.edu.tw}
\emailAdd{dns@astro.princeton.edu}

\abstract{Yes.  Future CMB experiments such as Advanced ACTPol and CMB-S4 should achieve
measurements with S/N of $> 0.1$ for the typical host halo of galaxies in redshift surveys.  These measurements
will provide complementary measurements of 
 the growth rate of large scale structure $f$ and the expansion rate of the Universe $H$
to galaxy clustering measurements.
This paper emphasizes that there is significant information in the anisotropy of
the relative pairwise kSZ measurements.
We expand the relative pairwise kSZ power spectrum in Legendre polynomials
and consider up to its octopole. Assuming that the noise in the filtered maps
is uncorrelated between the positions of galaxies in the survey,
we derive a simple analytic form for the power spectrum covariance of the relative pairwise kSZ temperature in redshift space.
While many previous studies have assumed optimistically that the optical depth of the galaxies $\tau_{\rm T}$ in the survey
is known, we marginalize over $\tau_{\rm T}$, to compute constraints 
on the growth rate $f$ and the expansion rate $H$.  
For realistic survey parameters, we find that combining kSZ and galaxy redshift survey data
reduces the marginalized $1$-$\sigma$ errors on $H$ and $f$ to $\sim 50\mathchar`-70\%$ compared to the galaxy-only analysis.}


\arxivnumber{arXiv:1606.06367}

\notoc
\maketitle
\flushbottom

\section{Introduction}

There is growing interest in using kinematic Sunyaev-Zel'dovich~\cite{Sunyaev1970kSZ,Sunyaev1972,Sunyaev1980,Ostriker1986} (hereafter, kSZ)
measurements as a cosmological probe.
The kSZ effect induces brightness temperature anisotropies in the Cosmic Microwave Background (CMB) radiation
due to the Doppler shift when the CMB photons scatter off a cloud of moving free electrons inside and around a galaxy cluster.
The temperature shift is sourced by the free electron peculiar velocity along the line-of-sight (LOS) with respect to the CMB rest frame.
Therefore, the kSZ effect offers a unique and powerful observational tool for measuring peculiar motions of galaxy clusters
on cosmologically interesting scales.
Measurements of the peculiar motions through the kSZ effect would be very valuable, since it
would not only allow us to test the predictions of the standard model, 
but also provide additional constraints on cosmological parameters, especially on 
the equation of state of dark energy and modified gravity
~\cite{DeDeo2005,Hernandez2006,Bhattacharya2007ApJ...659L..83B,Bhattacharya2008, Kosowsky2009PhRvD..80f2003K, Keisler2013,Ma2014,Mueller2015a,Alonso2016},
the Copernican Principle and the homogeneity of the Universe~\cite{Zhang2011,Planck2014XIII}, and neutrino mass~\cite{Mueller2015b}.

In addition to constraints on cosmological parameters of interest, 
the kSZ effect is also a potentially powerful probe to the ``missing baryon'' problem
~\cite{DeDeo2005,Bregman2007ARA&A..45..221B,Hernandez2008,Ho2009b,Hernandez2009,Shao2011,Hernandez2015,Schaan2016}
and galaxy formation feedback~\cite{Flender2016}.
Most of the baryons at low redshift ($z<2$) are thought to be in the warm hot intergalactic medium 
with temperature in a range of $10^5$-$10^7$ $K$~\cite{Cen2006}.
This hot diffuse ionized gas is very difficult to observe through its emission, 
because it is neither hot enough ($T < 10^{8}$ $K$) to be seen in $X$-ray observations, 
nor cold enough ($T > 10^3$ $K$) to be made into stars and galaxies.
Since the kSZ signal is directly proportional to the gas density and is irrelevant to the gas temperature,
it is sensitive to the ionized gas in the Universe.

With these motivations, there have been various efforts to detect the kSZ effect over four decades after the kSZ effect was first proposed 
in 1970~\cite{Kashlinsky2008,Kashlinsky2010,Kashlinsky2011ApJ...732....1K,Keisler2009,Osborne2011,Mody2012,Feindt2013}.
Early attempts yielded gradually improving  upper limits~\cite{Holzapfel1997,Mauskopf2000,Benson2003,Zemcov2003,Mauskopf2012,Lavaux2013}.
In 2012, Hand et al.~\cite{Hand2012} reported the first clear detection of the kSZ effect 
through the LOS pairwise velocity estimator~\cite{Ferreira1999} 
using CMB data from the Atacama Cosmology Telescope (ACT)~\cite{Swetz2011} with galaxy positions from the Baryon Oscillation Spectroscopic Survey (BOSS)
Data Release (DR) 9, which is a part of the Sloan Digital Sky Survey III (SDSSIII)~\cite{Eisenstein2011}.
In 2015, Planck collaboration~\cite{Planck2016A&A...586A.140P} has also detected the kSZ effect through
the same estimator as~\cite{Hand2012} and the kSZ temperature-reconstructed velocity field correlation
using galaxy positions from SDSS DR7~\cite{Abazajian2009ApJS..182..543A}.
Soergel et al.\cite{Soergel2016} has measured the pairwise estimator and detected the kSZ effect using 
CMB data from the South Pole Telescope (SPT)~\cite{George2015ApJ...799..177G} with cluster positions estimated by photometric redshift data from the Dark Energy Survey (DES)
~\cite{DES2005astro.ph.10346T,DES2016MNRAS.460.1270D}.
Schaan et al.~\cite{Schaan2016} has reported a detection of the kSZ signal 
by stacking the CMB temperature at the location of each halo, weighted by the corresponding reconstructed velocity,
using CMB data from ACTPol~\cite{Naess2014JCAP...10..007N} with galaxy positions from BOSS.
In addition, Sayers et al.~\cite{Sayers2013} has provided the claimed detection of the kSZ effect in a single source.
Hill et al.~\cite{Hill2016} and Ferraro et al.~\cite{Ferraro2016arXiv160502722F} have detected the kSZ effect through a three point correlation 
using CMB data from Planck with galaxy density distributions from the Wide-field Infrared Survey Explorer (WISE)~\cite{Wright2010AJ....140.1868W}.

The mean relative peculiar velocity averaged over pairs at separation $r$, so-called {\it mean pairwise velocity},
$v_{\rm pair}(r)$ was first introduced in the context of the BBGKY theory~\cite{Peebles1976Ap&SS..45....3P,Davis1977,Peebles1980lssu.book.....P}
and then in the fluid limit~\cite{Fisher1994MNRAS.267..927F,Juszkiewicz1998ApJ...504L...1J,Juszkiewicz1999ApJ...518L..25J},
and was measured by~\cite{Hand2012,Planck2016A&A...586A.140P,Soergel2016} through the kSZ effect at cosmological distances.
There are two issues in measuring the mean pairwise velocity as follows. One is 
that we observe only the line-of-sight component of the peculiar velocity rather than the full three-dimensional velocity~\cite{Ferreira1999}.
Another is that observed three-dimensional positions of galaxies 
are distorted along the LOS by redshift space distortions (RSD)~\cite{Sargent1977ApJ...212L...3S,Kaiser1987MNRAS.227....1K,Hamilton1998ASSL..231..185H}.
Thus, it is not possible to measure the mean pairwise velocity $v_{\rm pair}(r)$ as a function of the true distance, $r$, directly.
To interpret the kSZ measurements~\cite{Hand2012,Planck2016A&A...586A.140P,Soergel2016}
and the measurements of the bulk flow of the local Universe~\cite{Feldman2003,Feldman2010MNRAS.407.2328F} correctly,
we have developed the theoretical model of 
the LOS pairwise velocity in redshift space
and have investigated its non-linear properties in both configuration and Fourier spaces~\cite{Okumura2014JCAP...05..003O,Sugiyama2015arXiv150908232S},
where we have mainly focused on the statistics of the number density-weighted velocity which is straightforwardly obtained from kSZ (galaxy velocity)
data and simulations.

The mean pairwise velocity is related to the two-point correlation function of density fluctuations
by the pair conservation equation and isotropic symmetry assumption~\cite{Peebles1976Ap&SS..45....3P,Davis1977,Peebles1980lssu.book.....P},
where the time derivative of the correlation function averaged over a sphere of radius $r$ corresponds to the mean pairwise velocity.
However, this relationship only holds for conserved fields in real space (in isotropic clustering). 
As an alternative approach, we have shown a new relation 
between the density and LOS density-weighted velocity fields in Fourier space,
which holds in redshift space (in anisotropic clustering) and also for any discrete object and enables us to extend familiar techniques 
to compute the density power spectrum to the LOS density-weighted velocity power spectrum~\cite{Sugiyama2015arXiv150908232S}. 
There, we have computed the power spectrum/correlation function of the LOS density-weighted velocity 
in redshift space in the context of the Lagrangian perturbation theory.

The amplitude of the kSZ effect is proportional to $\tau_{\rm T} f$ where $\tau_{\rm T}$ is the optical depth in
the tracer population and $f$ is the growth rate of structure.  Most papers that explored the use 
of kSZ measurements for cosmology assume that $\tau_{\rm T}$ is known and non-evolving, so use the
kSZ measurements to trace the evolution of $f(z)$.  This approach could be risky, as galaxy formation feedback
has significant effects on the large-scale distribution of the electrons and can alter the amplitude of the kSZ
signal by up to 50\% \cite{Schaan2016, Flender2016}.  In this paper, we will make use of the angular
power spectrum of the kSZ effect.   Two effects, redshift-space distortions and the Alcock-Paczynski (AP) test~\cite{Alcock1979Natur.281..358A}, each leave their distinctive mark on the kSZ angular power spectrum. By measuring the dipole and octopole of the kSZ power spectrum~\cite{Sugiyama2015arXiv150908232S},
we can detect the signature of these effects and infer cosmology from the joint analysis of the galaxy and kSZ power spectrum independently of our knowledge of $\tau_{\rm T}$.

The outline of this paper is as follows.
Section~\ref{Sec:kSZ} reviews the kSZ effect and an observable density-weighted kSZ field in redshift space.
Section~\ref{Sec:Velocity} provides analytic expressions of a density-weighted LOS velocity field proportional to the kSZ observables,
its power spectra/correlation functions, and covariances of the power spectra/correlation functions.
Section~\ref{Sec:Noise} describes the dominant noise contributions to the kSZ signal.
Section~\ref{Sec:Simulations} details how $N$-body simulations were generated.
Section~\ref{Sec:Results} computes the cumulative signal-to-noise ratio for the kSZ power spectrum 
and forecasts constraints on cosmological parameters using the Fisher matrix formalism.
We present our conclusions in Section~\ref{Sec:Conclusion}: 
the combination of kSZ measurements from the next generation ``Stage-4'' ground based CMB experiment (CMB-S4;~\cite{Abazajian2015APh....63...55A})
with observations from the Dark Energy Spectroscopic Instrument (DESI;~\cite{Levi2013arXiv1308.0847L}) will
reduce the marginalized $1\mathchar`-\sigma$ errors on $H$ and $f$ to $\sim 50\mathchar`-70\%$ compared to a galaxy-only analysis.
For comparison with this paper, Appendix~\ref{ap:conservation} shows the relation between two-point correlation function of density fluctuations and the mean pairwise velocity
by the pair conservation equation and isotropic symmetry assumption~\cite{Peebles1976Ap&SS..45....3P,Davis1977,Peebles1980lssu.book.....P}.
Appendix~\ref{ap:derivative} gives detailed derivations of equations for a Fisher matrix analysis.

\section{Kinematic Sunyaev-Zel'dovich effect}
\label{Sec:kSZ}

The scattering of CMB photons with moving electrons induces brightness temperature anisotropies in CMB. 
This bulk motion-induced thermal distortion is called the kinematic Sunyaev-Zel'dovich effect
~\cite{Sunyaev1970kSZ,Sunyaev1972,Sunyaev1980,Ostriker1986},
\begin{eqnarray}
	  \delta T_{\rm kSZ}(\hat{n})= - T_{0} \int dl \sigma_{\rm T} n_{\rm e} \left( \frac{\vv_{\rm e} \cdot \hat{n}}{c} \right)
\end{eqnarray}
where $T_0$ is the average CMB temperature, 
$n_{\rm e}$ and $\vv_{\rm e}$ respectively denote the physical free electron number density and peculiar velocity,
$\sigma_{\rm T}$ is the Thomson scattering cross-section, and $c$ is the speed of light.
In the above expression, the integral $\int dl n_{\rm e}$ is performed along the LOS given by $\hat{n}$.
Since after recombination baryon fluctuations can quickly catch up dark matter fluctuations,
we assume that the peculiar velocity of free electrons is regarded as velocities of dark matter particles or dark halos: $\vv = \vv_{\rm e}$.
Assuming that the CMB photons scatter off only one cloud of moving free electrons until they reach observers,
the kSZ temperature at object~$i$ is proportional to the LOS peculiar velocity of object~$i$
\begin{eqnarray}
	  \delta T_i \simeq  \left( - T_{0}\frac{\tau_i }{c}  \right)  \vv_i \cdot\hat{n}_i,
	  \label{kSZ}
\end{eqnarray}
where $\vv_i$,  $\hat{n}_i$, and $\tau_i$ are the peculiar velocity, angular position, and Thomson optical depth of the $i$-th object, respectively.
The kSZ temperature $\delta T_i$ at object~$i$  is 
obtained with an aperture photometry filter~\cite{Planck2016A&A...586A.140P,Schaan2016} and a matched filter~\cite{Hand2011,Hand2012}.
The $\tau_i$ value depends on mass of the object ,$M$, and is proportional to $M$ in a simple model.
Since some fraction of the electrons in objects resides in the neutral medium
and does not take part in the Thomson scattering, 
the kSZ signal is proportional to the fraction of free electrons compared to the expected cosmological abundance~\cite{Schaan2016}.
Due to baryonic physics like star formation and feedback, 
the model in Eq.~(\ref{kSZ}) tends to overestimate the true kSZ signal
and the actual pairwise kSZ signal may be suppressed by $\sim50\%$. 
However, uncertainty from baryonic effects will affect only the global normalization of the pairwise kSZ signal,
not the shape as a function of scale, and therefore, 
the uncertainty will not affect optical depth estimates from kSZ measurements when we constrain the optical depth as a free parameter
~\cite{Flender2016}.
Throughout this paper, we assume that the optical depth at each object is the same $\tau_i = \tau_{\rm T}$,
where $\tau_{\rm T}$ is interpreted as an ``effective'' optical depth that is a proportionality constant to fit 
the observed kSZ temperature to the theoretical prediction of the LOS velocity.
This assumption corresponds to the assumption that there is no strong correlation between the optical depth $\tau$ and
the peculiar velocity $\vv$ of a given halo.
The value of $\tau_{\rm T}$ is expected to scale with the mass of the tracer population
and the aperture used in the analysis. 
For example, the Planck team measured $\tau_{\rm T} = \left( 1.4\pm0.5 \right)\times10^{-4}$~\cite{Planck2016A&A...586A.140P}
with an $8$ arcmin aperture for halos centered on the SDSS central galaxy catalogue.
The DESxSPT analysis~\cite{Soergel2016} measures the optical depth $\tau_{\rm T} = \left( 3.75\pm0.89 \right)\times10^{-3}$
using a matched filter~\cite{Li2014}.
The optical depth value from the DESxSPT analysis is significantly larger than the Planck result,
because the typical host halo masses used in the Planck analysis~\cite{Planck2016A&A...586A.140P}
are significantly below the mass range used in the DESxSPT analysis.
Furthermore, the matched filter optimized for the observed electron distribution
would provide higher effective optical depth than the Planck result.

Using Eq.~(\ref{kSZ}), we define a number density-weighted three-dimensional kSZ field:
\begin{eqnarray}
	  \delta T(\sr) = \sum_{i = 0}^{N - 1} \delta T_{i} \delta_{\rm D}\left( \sr - \sr_i \right),
	  \label{kSZfield}
\end{eqnarray}
where $\sr_i$ denotes the coordinate of object~$i$,
$\delta_{\rm D}$ represents the Dirac delta function, $N$ is the total number of observed objects.
It is important to stress here that the kSZ temperature is measured at the position of objects in redshift surveys.
For computational convenience,
we operate the flat sky limit and use the distant observer approximation $\hat{n} = \hat{n}_i$.
Then, the object position including RSDs is given by
\begin{eqnarray}
	  \sr_i = \xx_i + \frac{\vv_i\cdot\hat{n}}{aH}\hat{n},
	  \label{RSDcoordinate}
\end{eqnarray}
where $\xx_i$ is the position at object $i$ in real space, $H$ denotes the Hubble parameter, and $a$ is the scale factor.
Finally, from Eqs.~(\ref{kSZ}), (\ref{kSZfield}), and (\ref{RSDcoordinate}) the density-weighted kSZ field is represented by
\begin{eqnarray}
	  \delta T(\sr) = \left( -\frac{T_0 \tau_{\rm T}}{c} \right)\sum_{i=0}^{N-1} 
	  \left[\vv_i\cdot\hat{n} \right] \delta_{\rm D}\left( \sr - \xx_i - \frac{\vv_i\cdot\hat{n}}{aH}\hat{n} \right).
	  \label{T_s}
\end{eqnarray}

\section{Formalisms of LOS velocity fields}
\label{Sec:Velocity}

The number density-weighted kSZ field in Eq.~(\ref{T_s}) is proportional to the number density-weighted LOS velocity field.
In this section, we provide general formalisms of the number density-weighted LOS velocity field in redshift space.

Conventionally, the pairwise velocity statistics are computed by the so called ``pair-weighted'' average,
which is spatial (ensemble) averaging with the weighting factor
$\rho_1\rho_2/\langle \rho_1\rho_2\rangle$, where $\rho_1$ and $\rho_2$ are number density fields at points $\xx_1$ and $\xx_2$
\footnote{
In this paper, we do not use $n$ but $\rho$ to represent the ``number density'' in order to avoid confusion with the LOS direction $\hat{n}$.
}, and
the denominator $\langle \rho_1\rho_2\rangle$ yields the two-point correlation function of density fluctuations.
Since the pair-weighted average is not well-defined in an empty region $\rho(\xx)=0$~\cite{Zhang2015PhRvD..91d3522Z,Yu2015PhRvD..92h3527Y},
in this paper we replace the denominator of the pair-weighted average by the square of mean density fields $\bar{\rho}^2$
(or the two-point correlation function measured from a random catalog) which is scale-independent,
and employ the following weighting factor $\rho_1\rho_2/\bar{\rho}^2$.
(see also Appendix~B in~\cite{Okumura2014JCAP...05..003O}).
It is then not necessary to compute the two-point correlation function 
when we compare theoretical models with observations for the density-weighted statistics.
Thus, we focus on the number density-weighted velocity, which is defined as
\footnote{
The mass (number) density-weighted peculiar velocity field divided by the mean mass (number) density
is referred to as the ``momentum field''~\cite{Ma2002,Ma2014}, given by $\left( 1 + \delta \right) \vv$
in real space with $\delta$ being the density contrast.
}
\begin{eqnarray}
	  \pp(\xx) = \rho(\xx) \vv(\xx) = \sum_{i=0}^{N-1} \vv_i \delta_{\rm D}\left( \xx - \xx_i \right)
\end{eqnarray}
in real space, where $\rho(\xx)$ and $\vv(\xx)$ denote the number density and peculiar velocity at position $\xx$,
and $\xx_i$ and $\vv_i$ represent the position and peculiar velocity at object $i$.
For brevity we abbreviate the number density-weighted kSZ and LOS velocity fields to the kSZ and LOS velocity fields in what follows.

\subsection{LOS velocity fields}

We define the $n$-th moment of the LOS velocity field in redshift space as~\cite{Sugiyama2015arXiv150908232S}
\begin{eqnarray}
	  p_{\rm s}^{(n)}(\sr) &=& 
	  \sum_{i=0}^{N-1}\left[\vv_i\cdot  \hat{n}\right]^n\delta_{\rm D}\left( \sr - \xx_i - \frac{\vv_i\cdot\hat{n}}{aH}\hat{n}\right),
	  \label{mom}
\end{eqnarray}
where the subscript ``s'' denotes a quantity defined in redshift space.
The lowest order of $p_{\rm s}^{(n)}$ corresponds to the normal number density field in redshift space $p^{(0)}_{\rm s} = \rho_{\rm s}$.
The next order is proportional to the kSZ field in Eq.~(\ref{T_s})
\begin{eqnarray}
	\delta T(\sr) = - \left( \frac{T_0 \tau_{\rm T}}{c} \right) p_{\rm s}^{(1)}(\sr).
\end{eqnarray}
The volume average of the LOS velocity field is given by
\begin{eqnarray}
	  \bar{p}^{(n)}_{\rm s} = \frac{1}{V} \int d^3s p^{(n)}_{\rm s}(\sr)
	  = \bar{n} \left(  \frac{1}{N} \sum_{i=0}^{N-1} \left[ \vv_i\cdot\hat{n} \right]^{n}  \right),
	  \label{mean}
\end{eqnarray}
where $\bar{n}$ is the mean number density $\bar{n}=N/V$ with $V$ being a survey (simulation) volume.

Following~\cite{Sugiyama2015arXiv150908232S},
the Fourier transform of $p_{\rm s}^{(n)}$, 
defined as $p_{\rm s}^{(n)}(\kk) \equiv \int d^3s e^{-i\kk\cdot\sr} p_{\rm s}^{(n)}(\sr)$,
enables us to show a simple relation between $p_{\rm s}^{(n)}$ and $\rho_{\rm s}$,
\begin{eqnarray}
	  p_{\rm s}^{(n)}(\kk) = \left( i\frac{aH}{\kk\cdot\hat{n}} \right)^n \frac{d^n}{d\gamma^n} \rho_{\rm s}(\kk;\gamma)\bigg|_{\gamma = 1}.
	  \label{mom_den}
\end{eqnarray}
In the above expression, 
the density field in redshift space behaves as a generating function of the LOS velocity field, which is defined as
\begin{eqnarray}
  	\rho_{\rm s}(\kk;\gamma) =
	\sum_{i=0}^{N-1} e^{-i\kk\cdot\xx_i} e^{ -i\kk\cdot\hat{n} \frac{\vv_i\cdot\hat{n}}{aH} \gamma}.
\end{eqnarray}
The generating function for $\gamma = 1$ reduces to the normal density field:
$\rho_{\rm s}(\kk;\gamma = 1) = \rho_{\rm s}(\kk)$.
Provided that the velocity field $\vv$ is proportional to the linear growth rate $f= d \ln D/ d\ln a$
with $D$ being the linear growth factor, we derive
\begin{eqnarray}
	  p_{\rm s}^{(n)}(\kk) = \left( i\frac{aHf}{\kk\cdot\hat{n}} \right)^n \frac{\partial^n}{\partial f^n} \rho_{\rm s}(\kk).
	\label{rho}
\end{eqnarray}
The expressions in Eqs.~(\ref{mom_den}) and (\ref{rho}) relates the $n$-th moment of the LOS velocity field in redshift space
to the number density field including RSDs,
and can be used to compute $p_{\rm s}^{(n)}$ using the analytical expression of $\rho_{\rm s}$.

Assuming the conservation of the number of objects, $N={\rm const.}$,
the time-derivative of the density field in real space leads to the continuity equation,
which relates the time-derivative of the density field to the divergence of the density-weighted velocity:
$\dot{\rho} + \nabla\cdot\pp=0$ (see Appendix~\ref{ap:conservation}).
On the other hand, we stress here that
Eqs.~(\ref{mom_den}) and (\ref{rho}) hold without the conservation of the number of objects in redshift space.
Therefore, Eqs.~(\ref{mom_den}) and (\ref{rho}) can be applied to any discrete object (e.g., dark matter particles, halos, galaxies, and galaxy clusters)
even if the total number of objects may be time-dependent, and the galaxy clustering is anisotropic due to RSDs.

\subsection{Power spectrum and correlation function}
\label{Sec:Power}

From Eq.~(\ref{mom}) the estimator of a LOS velocity power spectrum, which can be used in simulations, is given by
\begin{eqnarray}
	  \widehat{P}_{\rm s}^{(n)(m)}(\kk)
	  &=&  \frac{V}{N^2} \left[ p_{\rm s}^{(n)}(\kk) \right]\left[ p_{\rm s}^{(m)}(\kk) \right]^* \nonumber \\
	  &=& \frac{V}{N^2}\sum_{i,j} \left[\vv_i\cdot \hat{n} \right]^n \left[ \vv_j\cdot\hat{n} \right]^m e^{-i\kk\cdot(\sr_i - \sr_j)},
	  \label{P_nm1}
\end{eqnarray}
where $\kk \neq 0$.
The shape of the power spectrum should be unaffected by the average of the LOS velocity field which is measured by the $\kk=0$ mode,
and $\widehat{P}^{(n)(m)}_{\rm s}(\kk=0) = 0$.
Note that the weighting factor $\rho_1\rho_2/\bar{\rho}^2$ is used in the above expression.
Similar to the LOS velocity field, 
the estimator of the LOS velocity power spectrum $\widehat{P}_{\rm s}^{(n)(m)}$ is derived from a generating function~\cite{Sugiyama2015arXiv150908232S}
\begin{eqnarray}
	  \widehat{P}_{\rm s}^{(n)(m)}(\kk)
	  &=& \left( -1 \right)^m \left( i\frac{aH}{\kk\cdot\hat{n}} \right)^{n+m}
		\frac{d^n}{d\gamma_1^n}\frac{d^m}{d\gamma_2^m}
		\widehat{P}_{\rm s}\left( \kk;\gamma_1,\gamma_2 \right)\bigg|_{\gamma_1 = \gamma_2 = 1},
	  \label{P_nm}
\end{eqnarray}
where the generating function is defined as
\begin{eqnarray}
	  \widehat{P}_{\rm s}\left( \kk;\gamma_1,\gamma_2 \right) =
	  \frac{V}{N^2} \rho_{\rm s}(\kk;\gamma_1) \rho^*_{\rm s}(\kk;\gamma_2),
	  \label{P_d_gf}
\end{eqnarray}
which reduces to the estimator of the density power spectrum in redshift space for $\gamma_1 = \gamma_2 = 1$:
$\widehat{P}_{\rm s}(\kk;\gamma_1=1,\gamma_2=1) = \widehat{P}_{\rm s}(\kk)$.
Furthermore, the estimator of the $n$-th moment of the relative pairwise LOS velocity power spectrum $\widehat{P}^{(n)}_{\rm s}$ is given by
\begin{eqnarray}
	  \widehat{P}^{(n)}_{\rm s}(\kk)
	   &=& \sum_{m=0}^{n} \frac{(-1)^m n!}{m!(n-m)!} \widehat{P}_{\rm s}^{(n-m)(m)}(\kk) \nonumber \\
	  &=&  \frac{V}{N^2}\sum_{i,j}
	  \left[ \vv_i\cdot\hat{n} - \vv_j\cdot\hat{n} \right]^n e^{-i\kk\cdot\left( \sr_i - \sr_j \right)} \nonumber \\
	  &=& \left( i\frac{aH}{\kk\cdot\hat{n}} \right)^n  \frac{d^n}{d\gamma^n}
	  \widehat{P}_{\rm s}\left( \kk;\gamma\right) \bigg|_{\gamma = 1},
	  \label{P_n}
\end{eqnarray}
where $\widehat{P}_{\rm s}(\kk;\gamma)$ is the generating function of $\widehat{P}^{(n)}_{\rm s}(\kk;\gamma)$, given by
\begin{eqnarray}
	\widehat{P}_{\rm s}(\kk;\gamma) = \widehat{P}_{\rm s}(\kk;\gamma,\gamma).
\end{eqnarray}
Similar to Eq.~(\ref{rho}),
provided $\vv \propto f$, we obtain~\cite{Sugiyama2015arXiv150908232S}
\begin{eqnarray}
	   \widehat{P}^{(n)}_{\rm s}(\kk) =  \left( i\frac{aHf}{\kk\cdot\hat{n}} \right)^n  \frac{\partial^n}{\partial f^n}
	  \widehat{P}_{\rm s}(\kk).
	  \label{P_n2}
\end{eqnarray}
These expressions in Eqs.~(\ref{P_nm}), (\ref{P_n}), and (\ref{P_n2}) hold for any discrete objects 
and relate the density power spectrum to the LOS velocity power spectra in redshift space
without the assumption of the pair conservation (see also Appendix~\ref{ap:conservation}).
The power spectra are derived from the ensemble average of their estimators:
$P_{\rm s}^{(n)(m)} = \langle \widehat{P}_{\rm s}^{(n)(m)}\rangle$ and $P_{\rm s}^{(n)} = \langle \widehat{P}_{\rm s}^{(n)}\rangle$.

In measuring $P_{\rm s}^{(n)(m)}$ from Eq.~(\ref{P_nm1}), 
the discreteness effect introduces a constant term, so-called the ``shot-noise'' term.
The LOS velocity power spectrum without the shot-noise term is given by
\begin{eqnarray}
	  \widetilde{P}^{(n)(m)}_{\rm s}(\kk) =  P^{(n)(m)}_{\rm s}(\kk) - \frac{1}{\bar{n}} \sigma_{\rm v}^{(n+m)},
	  \label{PwnS}
\end{eqnarray}
where
\begin{eqnarray}
	  \sigma_{\rm v}^{(n+m)} \equiv \left\langle   \frac{1}{N}\sum_{i=0}^{N-1} \left[ \vv_i\cdot\hat{n} \right]^{n+m} \right\rangle.
	  \label{LOSvelocity}
\end{eqnarray}
It should be noted that the LOS pairwise velocity power spectrum $P_{\rm s}^{(n\geq 1)}$ has no shot-noise term
due to the relative velocity weighting factor $\left[\vv_i\cdot \hat{n} - \vv_j\cdot\hat{n} \right]^n$.
As shown later in Sec.~\ref{Sec:Covariance}, the covariance matrix of $P^{(n)}$ has contributions from shot-noise terms.

The LOS velocity power spectra, even in real space, have the angular-dependence.
Statistically, these anisotropies are axially symmetric around the LOS in the distant-observer limit.
Then, it is common to expand the power spectra in Legendre polynomials $\LL_{\ell}(\mu)$,
where $\mu$ is the cosine of the angle to the LOS $\mu \equiv \hat{k}\cdot\hat{n}$, 
and average the power spectra around the LOS to give multi-pole moments,
\begin{eqnarray}
	  P^{(n)(m)}_{\ell}(k) &=& \left\langle
	  \frac{2\ell+1}{4\pi}\int d\varphi \int d\mu {\cal L}_{\ell}(\mu)   \widehat{P}_{\rm s}^{(n)(m)}(\kk) \right\rangle, \nonumber \\
	  P^{(n)}_{\ell}(k) &=& \left\langle \frac{2\ell+1}{4\pi}\int d\varphi \int d\mu {\cal L}_{\ell}(\mu)  
	  \widehat{P}_{\rm s}^{(n)}(\kk) \right\rangle,
	  \label{Pk_legendre}
\end{eqnarray}
where $\varphi$ denotes the rotation angle around the LOS.
The axial symmetry shows that the $P_{\ell}^{(n)(m)}$ for $n + m$ = even and $n + m$ = odd should be expanded by even- and odd-pole moments.
Similarly, the $P_{\ell}^{(n)}$ for $n$ = even and $n$ = odd only contain even- and odd-pole moments.

The inverse Fourier transform of the LOS velocity power spectra in Eq.~(\ref{Pk_legendre})
leads to the analytic forms of LOS velocity correlation functions
\begin{eqnarray}
	  \xi_{\ell}^{(n)(m)}(s) &=&  i^{\ell} \int \frac{dk k^2}{2\pi^2} j_{\ell}(ks) P_{\ell}^{(n)(m)}(k), \nonumber \\
	  \xi_{\ell}^{(n)}(s) &=&  i^{\ell} \int \frac{dk k^2}{2\pi^2} j_{\ell}(ks) P_{\ell}^{(n)}(k).
	  \label{Xi_legendre}
\end{eqnarray}
where $j_{\ell}(ks)$ are the spherical Bessel functions of order $\ell$.
The dipole of the first moment of the LOS relative pairwise velocity correlation function $\xi_{\ell=1}^{(1)}(s)$ is closely related to
the numerator of the mean pairwise velocity $v_{\rm pair}(r)$.
The $\xi_{\ell=1}^{(1)}(s)$ computes the LOS peculiar velocity in redshift space,
while the $v_{\rm pair}(r)$ computes the full three-dimensional peculiar velocity in real space.
We follow the convention that 
if two galaxy clusters are moving toward each other, their contribution to $\xi_{\ell=1}^{(1)}(s)$ is negative $\xi_{\ell=1}^{(1)}(s) < 0$,
and if moving apart, positive $\xi_{\ell=1}^{(1)}(s)>0$.
Gravitational attraction predicts a slight tendency of any pair of galaxy clusters 
to be moving toward rather than away from each other at large scales, resulting in the negative value of $\xi_{\ell=1}^{(1)}(s)$.  
The linear RSD effect, known as the Kaiser effect, increases the amplitude of the pairwise velocity dipole at large scales.
On the other hand, at non-linear small scales
the RSD effect displaces the galaxy clusters away from each other, and in redshift space the sign of $\xi_{\ell=1}^{(1)}(s)$ thus changes around $30$ $h^{-1}{\rm Mpc}$ from negative to positive~\cite{Okumura2014JCAP...05..003O,Sugiyama2015arXiv150908232S},
which can be interpreted in analogy with the Finger-of-God effect in the quadrupole moment of the density power spectrum.

\subsection{Geometric distortions}

All LOS velocity power spectra are measured relative to a fiducial cosmology.
The difference between the fiducial and true values of an angular diameter distance $D_{\rm A}(z)$ and the Hubble parameter $H(z)$
yields~\cite{Ballinger1996MNRAS.282..877B,Matsubara1996ApJ...470L...1M,Blazek2014JCAP...04..001B}
\begin{eqnarray}
	  P^{(n)(m)}_{\rm obs}(k_{\rm fid}, \mu_{\rm fid}) &=&  \frac{1}{\alpha^3}P^{(n)(m)}_{\rm true}(k_{\rm true}, \mu_{\rm true}), \nonumber \\
	  k_{\rm true} &=&  k_{\rm fid}\frac{1+\varepsilon}{\alpha}
	  \left[ 1 + \mu_{\rm fid}^2\left( (1+\varepsilon)^{-6}-1 \right)	 \right]^{\frac{1}{2}}, \nonumber \\
	  \mu_{\rm true} &=&  \mu_{\rm fid}\frac{1}{(1+\varepsilon)^3}
	  \left[ 1 + \mu_{\rm fid}^2\left( (1+\varepsilon)^{-6}-1 \right)	 \right]^{-\frac{1}{2}},
	  \label{Geo1}
\end{eqnarray}
where ``obs'' stands for ``observed'' and ``fid'' stands for ``fiducial''.
Power spectrum measurements are labeled by $k_{\rm fid}$ and $\mu_{\rm fid}$,
which are held fixed under numerical derivations with respect to parameters.
As an intuitive parameterization of geometric distortions,
we employ isotropic dilation $\alpha$ and anisotropic warping $\varepsilon$ parameters, defined as~\cite{Padmanabhan2008PhRvD..77l3540P}
\begin{eqnarray}
	  \alpha &=& \left[ \frac{D_{\rm A}^2(z)}{D_{\rm A, fid}^2(z)}\frac{H_{\rm fid}(z)}{H(z)} \right]^{\frac{1}{3}}, \nonumber \\
	  1 + \varepsilon &=& \left[ \frac{D_{\rm A, fid}(z)}{D_{\rm A}(z)}\frac{H_{\rm fid}(z)}{H(z)} \right]^{\frac{1}{3}}.
	  \label{Geo2}
\end{eqnarray}
Note that if there is no isotropic shift, then $\alpha=1$. Similarly, the lack of anisotropy implies $\varepsilon=0$.

\subsection{Covariance matrix}
\label{Sec:Covariance}

We next turn to the covariance matrix between the LOS relative pairwise velocity power spectra,
which describes statistical uncertainties of the power spectrum measurement.
Using Eqs.~(\ref{P_n}) and (\ref{Pk_legendre}), the covariance matrix of the LOS relative pairwise velocity power spectrum
is derived from that of the density power spectrum by
\begin{eqnarray}
	  {\rm Cov}\left( \widehat{P}_{\ell_1}^{(n_1)}(k_1),  \widehat{P}_{\ell_2}^{(n_2)}(k_2)\right)
	  &=& \left( \frac{ 2\ell_1+1 }{4\pi}  \right)
	  \left( \frac{2\ell_2+1 }{4\pi}  \right)
	  \int d\mu_1 d\varphi_1
	   \int d\mu_2 d\varphi_2
	  {\cal L}_{\ell_1}(\mu_1)
	  {\cal L}_{\ell_2}(\mu_2) \nonumber \\
	  &\times& \left( i\frac{aH}{\kk_1\cdot\hat{n}} \right)^{n_1}
	  \left( i\frac{aH}{\kk_2\cdot\hat{n}} \right)^{n_2}
	  \frac{d^{n_1}}{d\gamma_1^{n_1}}
	  \frac{d^{n_2}}{d\gamma_2^{n_2}}
	   {\rm Cov}\left( \widehat{P}_{\rm s}(\kk_1;\gamma_1),  \widehat{P}_{\rm s}(\kk_2;\gamma_2) \right)
	  \Big|_{\gamma_1 = \gamma_2 = 1}. \nonumber \\
	  \label{Cov}
\end{eqnarray}  
The covariance between the power spectra
can be formally expressed in terms of unconnected and connected contributions in the cumulant expansion,
so-called the Gaussian and non-Gaussian terms, respectively~\cite{Meiksin1999MNRAS.308.1179M,Scoccimarro1999ApJ...527....1S}.
The Gaussian term has only diagonal elements of the covariance matrix, and the power spectrum estimates of different scales (bins) are uncorrelated.
On small scales,
non-vanishing off-diagonal elements arise from the non-Gaussian term which is represented by the trispectrum~\cite{Rimes2005MNRAS.360L..82R,Rimes2006MNRAS.371.1205R,Takahashi2009,Kiessling2011MNRAS.416.1045K}.
In this paper, we ignore these off-diagonal contributions, as these effects become
important on smaller scales than the kSZ noise contributions ($k\sim0.2\, h\, {\rm Mpc}^{-1}$) discussed in Sec.~\ref{Sec:Noise}.
Then, using Eq.~(\ref{Cov}) we obtain 
\begin{eqnarray}
	  {\rm Cov}\left( \widehat{P}_{\ell_1}^{(n_1)}(k_1),  \widehat{P}_{\ell_2}^{(n_2)}(k_2)\right) 
	  &=&  \frac{\delta^{\rm K}_{k_1k_2}}{N_{\rm mode}(k_1)} C_{\ell_1\ell_2}^{(n_1)(n_2)}(k_1)
	  \label{Cov_g}
\end{eqnarray}
where
\begin{eqnarray}
	  C^{(n_1)(n_2)}_{\ell_1 \ell_2}(k) 
	  &=&\frac{2\left( 2\ell_1+1 \right)\left( 2\ell_2+1 \right)}{4\pi}
	 \int d\varphi \int d\mu \LL_{\ell_1}(\mu) \LL_{\ell_2}(\mu) 
	  \nonumber \\
	 &\times&	
	 \left( i\frac{aH}{\kk\cdot\hat{n}} \right)^{n_1+n_2} 
	 \frac{d^{n_1}}{d\gamma_1^{n_1}}\frac{d^{n_2}}{d\gamma_2^{n_2}}
	 \left[  P_{\rm s}\left( \kk;\gamma_1,\gamma_2 \right) P_{\rm s}\left( \kk;\gamma_2,\gamma_1 \right)  \right]\bigg|_{\gamma_1 = \gamma_2 = 1}.
	 \label{Gaussian_C1}
\end{eqnarray}
In the above expression,
$N_{\rm mode}(k) = \frac{4\pi k^2\Delta k V}{(2\pi)^3}$
is the number of independent Fourier modes in a bin with a bin width $\Delta k$, and $\delta^{\rm K}$ denotes the Kronecker delta
defined such that $\delta^{\rm K}_{k_1k_2}=1$ if $k_1=k_2$ within the bin width, otherwise zero.

The simplest analytic form of the covariance between the LOS relative pairwise velocity correlation functions
assumes the Gaussian approximation, given by
\begin{eqnarray}
	  {\rm Cov}\left( \widehat{\xi}_{\ell_1}^{(n_1)}(s_1),  \widehat{\xi}_{\ell_2}^{(n_2)}(s_2)\right)
	  &=& 
	  \frac{i^{\ell_1+\ell_2}}{V}\int \frac{dk k^2}{2\pi^2} j_{\ell_1}(ks_1)j_{\ell_2}(ks_2) C_{\ell_1\ell_2}^{(n_1)(n_2)}(k),
\end{eqnarray}
where we used Eqs.~(\ref{Xi_legendre}) and (\ref{Gaussian_C1}).
Note that the covariance between the LOS relative pairwise velocity correlation function has off-diagonal elements even in the Gaussian approximation.

\section{Noise estimation}
\label{Sec:Noise}

For extracting the kSZ signal from data, 
the aperture photometry filter (the compensated top-hat filter) is applied at the position of each galaxy
with a characteristic filter scale $\theta_{\rm F}$~\cite{Planck2016A&A...586A.140P,Schaan2016}.
Applying the aperture photometry filter consists in taking the average CMB temperature within a disk of radius $\theta_{\rm F}$ 
and subtracting from it the average CMB temperature within an outer circular ring of equal area
to reduce the effect of the primary CMB fluctuations on scales larger than the filter width.
\cite{Schaan2016} has measured the kSZ effect at different filter scales 
by stacking the CMB temperature at the location of each halo, weighted by the corresponding reconstructed velocity,
and has found that the aperture radii of $\theta_{\rm F}=$ $2$-$3$ arcmin, 
which are slightly larger than a typical halo size of the CMASS sample $\sim1.4'$ for the ACTPol beam ${\rm FWHM}=1.4'$, has the best performance. 
Therefore, we focus on the filter scale of $\theta_{\rm F}=2$ arcmin in what follows.
The optimal filter width will depend on both the experimental setup (beam size and detector noise level) and the electron distribution around the target
galaxy population.

In the kSZ measurement, there are  two primary sources of noise: detector noise and primary CMB anisotropies.
For single frequency experiments, tSZ can also be a significant noise source.
A CMB instrument noise level in Advanced ACTPol~\cite{Calabrese:2014gwa,Henderson2015arXiv151002809H} is $\sim 8$ $\mu K$-arcmin
for a single-frequency measurement using the $150\ {\rm GHz}$ channel,
resulting in the detector noise of $\sim 3$ $\mu K$ on the kSZ signal for $2$ arcmin aperture.
CMB-S4~\cite{Abazajian2015APh....63...55A} aims to achieve  an instrument noise level of $\sim 1$ $\mu K$-arcmin,
and the detector noise on the kSZ temperature reduces to $\sim 0.4$ $\mu K$ for the $2$ arcmin aperture.
In practice one could improve the significance by combining frequency channels, and we expect higher significance detection.
The aperture photometry filtering should remove most of the primary CMB contribution to the kSZ temperature but unavoidably leave some residuals.
This residual primary temperature is estimated by Figure $6$ in~\cite{Hernandez2006},
and the temperature is $\sim 2$ $\mu K$ for $\theta_{\rm F} = 2$ arcmin.
Ultimately, a matched filter~\cite{Li2014} optimized for the observed electron distribution
would provide the highest S/N measurements of the  kSZ signal.

According to the above discussion, we describe observed kSZ temperature fluctuations as follows
\begin{eqnarray}
	  \delta T_i = \left( -\frac{T_0\tau_{\rm T}}{c}\right)\vv_i \cdot \hat{n} 
	  + \delta T_{{\rm N},i}
	  \label{NoiseModel}
\end{eqnarray}
where $\delta T_{ {\rm N},i}$ denotes the total noise at object~$i$ which includes all of the noise components for the kSZ measurement:
primary CMB anisotropies, detector noise, tSZ, and residual foregrounds.
Here, we assume that the residual noise at the cluster positions is an  uncorrelated, Gaussian
field that  satisfies
\begin{eqnarray}
	  \langle \delta T_{ {\rm N},i} \rangle_{\rm c} &=&  \mu_{\rm N}, \nonumber \\
	  \left\langle \delta T_{ {\rm N}, i} \delta T_{ {\rm N}, j} \right \rangle_{\rm c}
	  &=&  \sigma_{\rm N}^2 \delta_{ij},
	  \label{NoiseDispersions}
\end{eqnarray}
where $\langle \cdots \rangle_{\rm c}$ means the cumulants, and the cumulants beyond the first two are zero for the Gaussian distribution.
The presence of the tSZ effect will yield non-zero mean value $\mu_{\rm N} \neq 0$. 
For a properly designed experiment, the detector noise in the maps is close to white, so can
be treated as an uncorrected random field. 
Even though there are atmospheric noises that could introduce large-scale noise correlations,
the aperture photometry filter removes any large-scale background variations, so that they are no issue for kSZ measurements.
The level of residual foreground contamination should
also be uncorrelated between halos.  While the primary CMB in the raw maps has significant large-scale correlation,
the role of the aperture filter is to remove these correlations, so that the primary CMB signal in the {\it filtered}
map is nearly uncorrelated on scales larger than those of the compensated filter function~\cite{Schaan2016}.

Under the above assumptions about the noise properties,
the relative pairwise kSZ power spectrum is proportional to the first moment of the LOS relative pairwise velocity power spectrum 
(Eq.~(\ref{Pk_legendre}))
\begin{eqnarray}
	  P_{\rm kSZ, \ell}(k) =  -\left \langle \frac{2\ell+1}{4\pi} \int d\varphi \int d\mu \LL_{\ell}(\mu)\frac{V}{N^2} \sum_{i,j}
	  \left[ \delta T_i - \delta T_j \right] e^{-i\kk\cdot\left( \sr_i - \sr_j \right)} \right\rangle
	= \left( \frac{T_0 \tau_{\rm T}}{c} \right) P_{\ell}^{(1)}(k),
	\label{PkSZP}
\end{eqnarray}
which is unaffected by the noise.
The covariance matrix of the relative pairwise kSZ power spectrum is given by 
\begin{eqnarray}
	  {\rm Cov}\left(\widehat{P}_{\rm kSZ,\ell_1}(k_1), \widehat{P}_{\rm kSZ,\ell_2}(k_2)  \right)
	  &=&  \left( \frac{T_0 \tau_{\rm T}}{c} \right)^2
	  {\rm Cov}\left(\widehat{P}_{\ell_1}^{(1)}(k_1), \widehat{P}_{\ell_2}^{(1)}(k_2)  \right)  \nonumber \\
	  &+&
	  \frac{2\ell_1+1}{4\pi}\frac{2\ell_2+1}{4\pi}\int d\varphi_1 d\mu_1
	  \int d\varphi_2 d\mu_2 \LL_{\ell_1}(\mu_1)\LL_{\ell_2}(\mu_2) \nonumber \\
	  &&
	  \left(  \frac{V}{N^2}  \right)^2
	  \sum_{ijkl} \left\langle \left[ \delta T_{ {\rm N},i} - \delta T_{ {\rm N},j} \right]
	 \left[ \delta T_{ {\rm N},k} - \delta T_{ {\rm N},l} \right] \right\rangle 
	  \left\langle e^{-i\kk_1\cdot\left( \sr_i - \sr_j \right)}
	  e^{-i\kk_2\cdot\left( \sr_k - \sr_l \right)} \right\rangle, \nonumber \\
	 \label{CovNoise}
\end{eqnarray}
where we used the assumption that $\delta T_{ {\rm N}}$ is uncorrelated with positions and velocities of objects.
Using Eq.~(\ref{NoiseDispersions}),
the second term in Eq.~(\ref{CovNoise}), which represents the noise contribution to the covariance matrix, is given by
\begin{eqnarray}
&&	\frac{\delta^{{\rm K}}_{k_1 k_2}}{N_{\rm mode}(k_1)} \frac{2\left( 2\ell_1+1 \right)\left( 2\ell_2+1 \right)}{4\pi}
	 \int d\varphi d\mu \LL_{\ell_1}(\mu) \LL_{\ell_2}(\mu)  
	 \left[ - 2 \frac{\sigma_{\rm N}^2}{\bar{n}} P_{\rm s}(\kk)\right] \nonumber \\
	 &+& 	  \frac{2(2\ell_1+1)(2\ell_2+1)}{(4\pi)^2}\frac{1}{V}\int d\varphi_1  d\mu_1
	 d\varphi_2 d\mu_2 \LL_{\ell_1}(\mu_1)\LL_{\ell_2}(\mu_2)  \left[ 2\frac{\sigma_{\rm N}^2}{\bar{n}} B_{\rm s}(\kk_1,\kk_2)  \right],
\end{eqnarray}
where the first term is the Gaussian term, and 
the second term yields from the non-Gaussian term that is represented by the bispectrum $B_{\rm s}$ of the density field in redshift space, defined as
\begin{eqnarray}
	  V B_{\rm s}(\kk_1,\kk_2) = 
	  \left\langle \delta_{\rm s}(-\kk_1-\kk_2) \delta_{\rm s}(\kk_1)\delta_{\rm s}(\kk_2) \right \rangle.
\end{eqnarray}
In this paper, we ignore the above non-Gaussian term and leave the study of the effect as future works.
We note here that the mean value $\langle \delta T_{\rm N}\rangle = \mu_{\rm N}$ does not appear in the covariance matrix expression, 
which means that tSZ contamination is likely not a dominant issue under the noise assumption in Eq.~(\ref{NoiseDispersions}).

For computational convenience, we define the following inverse signal-to-noise ratio
\begin{eqnarray}
	  R_{\rm N} = \frac{\sigma_{\rm N}}{\sigma_{\rm kSZ}}
\end{eqnarray}
where $\sigma_{\rm kSZ}$ is a typical kSZ temperature estimate, defined as
\begin{eqnarray}
	  \sigma_{\rm kSZ} \equiv \left( \frac{T_0 \tau_{\rm T}}{c} \right) \sqrt{\sigma_{\rm v}^{(2)}}
\end{eqnarray}
with $(\sigma_{\rm v}^{(2)})^{1/2}$ being the LOS velocity dispersion in Eq.~(\ref{LOSvelocity}).
Finally, using Eqs.~(\ref{PwnS}), (\ref{Gaussian_C1}), (\ref{CovNoise}), and (\ref{NoiseDispersions}),
we derive the covariance matrix of the relative pairwise kSZ power spectrum in the Gaussian approximation as follows:
\begin{eqnarray}
	 && {\rm Cov}\left( \widehat{P}_{\rm kSZ, \ell_1}(k_1), \widehat{P}_{\rm kSZ, \ell_2}(k_2) \right)
	  = \frac{\delta^{{\rm K}}_{k_1 k_2}}{N_{\rm mode}(k_1)} C_{\ell_1\ell_2}(k_1), \nonumber \\
	  && C_{\ell_1\ell_2}(k)
	  =  \frac{2\left( 2\ell_1+1 \right)\left( 2\ell_2+1 \right)}{4\pi}
	 \int d\varphi \int d\mu \LL_{\ell_1}(\mu) \LL_{\ell_2}(\mu)  \nonumber \\
	 &&\hspace{1.3cm}\times \left( \frac{T_0\tau_{\rm T}}{c} \right)^2
	 \Bigg[2 \left(  \widetilde{P}_{\rm s}^{(1)(0)}(\kk)\right)^2 - 
	 2 \left(  \widetilde{P}_{\rm s}^{(1)(1)}(\kk) 
	 + \left( 1 + R_{\rm N}^2 \right) \frac{\sigma_{\rm v}^{(2)}}{\bar{n}}  \right)
	 \left( \widetilde{P}_{\rm s}(\kk)  + \frac{1}{\bar{n}} \right) \Bigg],
	 \label{kSZCov}
\end{eqnarray}
where we used $ \langle \sigma^{(1)}_{\rm v}\rangle = 0$. Throughout this paper, we adopt this simple noise model, and defer a more detailed noise treatment, e.g. tSZ contamination and non-Gaussian terms, for future studies. In particular, the uncorrelated Gaussian noise assumption in Eq.~(\ref{NoiseDispersions}) is technically not fully satisfied for the tSZ contribution to the noise; therefore, we need a further extension of our noise model or should use a CMB map free of tSZ constructed from multiple frequency data.

In~\cite{SugiyamaInPrep}, we measure the pairwise kSZ power spectrum from a CMB map free of tSZ that jointly uses both Planck~\cite{Planck2016A&A...594A..13P} and WMAP9~\cite{Bennett2013ApJS..208...20B} data, so-called WPR2~\cite{Bobin:2015fgb}, with the LOWZ and CMASS galaxy samples from BOSS DR12~\cite{Alam2015ApJS..219...12A}. There, we compare the standard deviations (the diagonal elements of the covariance matrix) computed from the noise model in Eq.~(\ref{kSZCov}) with internal error estimates from data itself using the jackknife~\cite{Quenouille1956,Tukey1958} and bootstrap~\cite{Efron1979} methods; we find a good agreement between the standard deviation estimates from the theory and the data. This fact validates our noise model in Eq.~(\ref{kSZCov}) and means that the uncorrelated Gaussian noise model in Eq.~(\ref{NoiseDispersions}) can explain dominant contributions to the covariance matrix as expected.

We construct a model for the optical depth in~\cite{SugiyamaInPrep} using the assumption that the projected electron profile is approximated by a Gaussian distribution~\cite{Schaan2016}. The Gaussian profile is a simplifying approximation and might be become inadequate for future high significance detections of the kSZ effect; therefore, it will be necessary to study other gas profiles such as the $\beta$-profile~\cite{Cavaliere1976A&A....49..137C} and the NFW profile~\cite{Navarro1996ApJ...462..563N} in future works.
Our optical depth model depends on the halo mass $M$, the aperture photometry radius $\theta_{\rm F}$, 
the effective beam full-width at half-maximum (FWHM) of CMB experiments, and redshift $z$.
Using the model, we estimate the optical depth values of $\tau_{\rm T}=5.1\times10^{-5}$ for Planck (${\rm FWHM}\sim 5'$)
and $\tau_{\rm T}=8.4\times10^{-5}$ for future CMB measurements such as Advanced ACTPol and CMB-S4 (${\rm FWHM}\sim 1'$),
where we assume a typical halo catalogue with the mean halo mass $M = 5.0\times10^{13}\, M_{\odot}$ at a redshift of $z=0.35$ that is similar to the LOWZ sample in BOSS.
Then, for the LOWZ-like halo catalogue (see Sec.~\ref{Sec:Simulations}) we estimate typical kSZ temperatures of $\sigma_{\rm kSZ} = 0.14\,\mu K$ for Planck 
and $\sigma_{\rm kSZ} = 0.24\, \mu K$ for CMB-S4.

As rough estimates, the rms noise in the aperture filtered maps for Advanced ACTPol and CMB-S4 will be  $\sigma_{\rm N}\sim$ $2$-$5$ $\mu K$
($R_{\rm N}\sim 10\mathchar`-30$) with a fixed aperture radius of $\theta_{\rm F}=2$ arcmin. 
For simplicity, we set the inverse signal-to-noise ratio as
\begin{eqnarray}
	R_{\rm N} =  0,\ 5,\ 10,\ {\rm and}\, 15
	  \label{NoiseRatio}
\end{eqnarray}
and investigate the impact of the noise levels  on cosmological information in the relative pairwise kSZ power spectrum.

\section{Simulations}
\label{Sec:Simulations}

To compute the relative pairwise kSZ power spectrum and its covariance, we use $N$-body simulations.
We have $48$ independent simulations generated with the cosmological $N$-body simulations code $Gadget2$~\cite{Springel:2000yr,Springel:2005mi}.
The simulations were all started from a redshift of $z=29$ 
with initial conditions in the 2LPT approximation generated by the $2LPT$ code~\cite{Crocce:2006ve}.
The fiducial cosmology corresponds to the best fitting parameters in the ${\rm Planck2015}$ data~\cite{Planck2016A&A...594A..13P}:
$\Omega_{\rm m} = 0.308$, $\Omega_{\rm \Lambda} = 0.692$,
$\Omega_{\rm b} = 0.048$, $h = 0.678$, $\sigma_8 = 0.815$, and $n_s = 0.9608$.
Linear matter power spectra are generated with  $CLASS$~\cite{Lesgourgues:2011re}.
There are $512^{3}$ particles in a box of $L = 1024\ h^{-1} {\rm Mpc}$
which leads to a particle mass of $m_{\rm p} = 6.8\times10^{11}\ h^{-1}M_{\odot}$.

To compute the LOS velocity power spectrum,
we use a fast Fourier transform (FFT) with an $N_{\rm grid} = 512$ grid on an axis.
Dark matter particles and halos in the $N$-body simulations
are distributed on the FFT grid using the triangular-shaped cloud (TSC), and 
the LOS velocity power spectrum is estimated at each $\kk$ by~\cite{Jing:2004fq}
\begin{eqnarray}
	  \widetilde{P}^{(n)(m)}_{\rm s}(\kk) &=&
	  \left\langle\frac{\frac{V}{N^2}
	  \left[ p_{\rm s, FFT}^{(n)}(\kk) \right]\left[ p_{\rm s, FFT}^{(m)}(\kk) \right]^* - P^{(n)(m)}_{\rm shot, TSC}(\kk)}
	  {W_{\rm TSC}^2(\kk)}  \right\rangle_{\rm sim=48}, \nonumber \\
	  \widehat{P}^{(n)(m)}_{\rm shot, TSC}(\kk) &=&
	\left(    \frac{1}{N}\sum_{i=0}^{N-1} \left[ \vv_i\cdot\hat{n} \right]^{n+m} \right)
	  \frac{1}{\bar{n}}\prod_{i=x,y,z}
	  \left[ 1 - \sin^2\left( \frac{\pi k_i}{2 k_{\rm N}} \right) + \frac{2}{15}\sin^4\left( \frac{\pi k_i}{2k_{\rm N}} \right) \right],
	  \nonumber \\
	  W_{\rm TSC}(\kk) &=&  \prod_{i=x,y,z}\left[ {\rm sinc}\left( \frac{\pi k_i}{2 k_{\rm N}} \right) \right]^3,
\end{eqnarray}
where $k_{\rm N} = \pi N_{\rm grid}/L$ is the Nyquist frequency,
and $\langle \cdots \rangle_{\rm sim=48}$ denotes the mean of 48 simulations.
We substitute the above 3-dimensional power spectrum into Eqs.~(\ref{P_n}) and (\ref{kSZCov})
to compute the relative pairwise kSZ power spectrum and its covariance.
We set the bin width as $\Delta k = 2\pi/L = 0.006\, h{\rm Mpc}^{-1}$,
which is similar to the bin width used in the galaxy power spectrum analysis in BOSS:
e.g., $\Delta k = 0.005\, h{\rm Mpc}^{-1}$~\cite{Beutler2014MNRAS.443.1065B} and $\Delta k = 0.01\, h{\rm Mpc}^{-1}$~\cite{Beutler2016arXiv160703150B}.

We assume that central galaxies exhibit the clustering properties similar to dark matter halos.
Since some galaxy types used in clustering measurements are primarily central objects (e.g., luminous red galaxies),
we compute the LOS relative pairwise velocity power spectrum for halos to evaluate the relative pairwise kSZ power spectrum using Eq.~(\ref{PkSZP}).
Similarly, we regard the halo density power spectrum as the galaxy power spectrum,
where baryonic effects will be negligible on the scales of interest ($k< 0.2\, h\, {\rm Mpc}^{-1}$)~\cite{Osato2015ApJ...806..186O}.
Dark matter halos are identified using the friends-of-friends algorithm~\cite{Davis1985ApJ...292..371D}
with a linking length of $0.2$ times the mean particle separation and at least $20$ particles.
Due to finite simulation snapshots, we focus on a result at $z=0.35$, even though
this redshift is somewhat lower than the effective redshift of the BOSS~\cite{Eisenstein2011} and DESI~\cite{Levi2013arXiv1308.0847L} surveys.
Since the kSZ and tSZ signals are comparable for the low-mass cluster ($M\sim10^{13}\ M_{\odot}$) which are far more abundant~\cite{Hand2012},
we employ a mass range of $10^{13}\ h^{-1}M_{\odot} < M < 10^{14}\ h^{-1}M_{\odot}$.
We fix the mean number density of halos to $\bar{n} = 3.0 \times 10^{-4}\ h^{3}{\rm Mpc}^{-3}$, which is similar to those assumed in BOSS and DESI,
by randomly selecting halos from each simulation.
Then, the linear bias value is $b=1.71$, which is computed by the ratio between the halo and matter power spectra 
in real space at $k=0.006\ h{\rm Mpc}^{-1}$.
These parameter choices are sufficient for our purpose, which 
investigates how the joint analysis of the galaxy and kSZ power spectra improve constrains on cosmological parameters 
compared to the galaxy-only analysis.
Since the volume of simulations is $V \sim1\ h^{-3}{\rm Gpc}^3$,
we can obtain rough error estimates of cosmological parameters for each survey in Sec.~\ref{Sec:Fisher}
by multiplying by $\sqrt{h^{-3}{\rm Gpc}^3/V_{\rm s}}$, where $V_{\rm s}$ is a survey volume
(e.g.,  $V_{\rm s} \sim 10.5\ h^{-3}{\rm Gpc}^3$ for DESI and $V_{\rm s}\sim3.5\ h^{-3}{\rm Gpc}^3$ for BOSS).

\begin{figure}[t]
	\centering
	\includegraphics[width=\columnwidth]{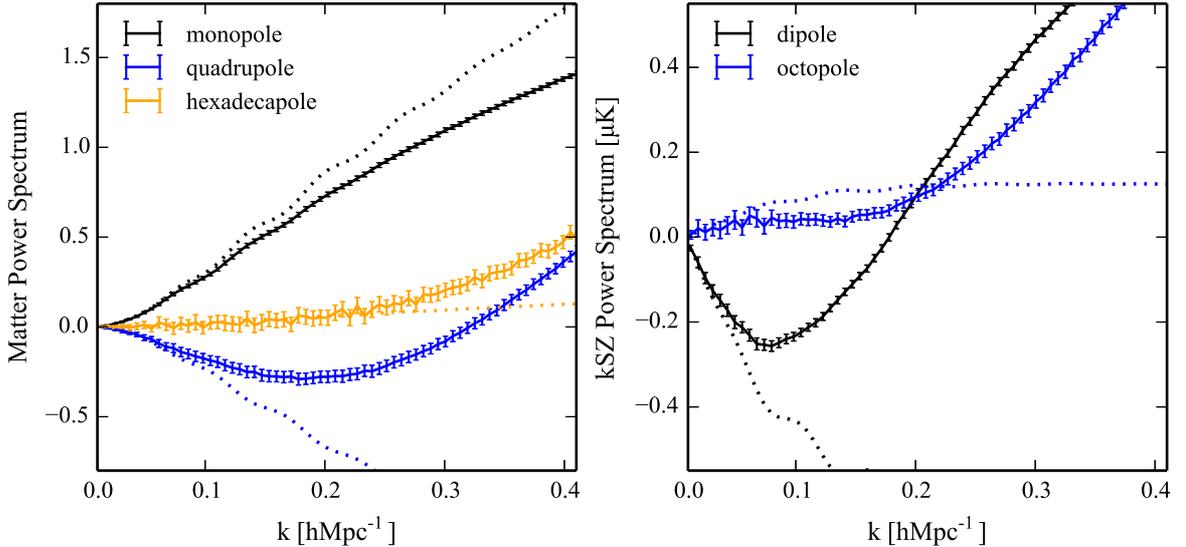}
	\caption{
	Multi-pole moments of the galaxy (left) and relative pairwise kSZ (right) power spectra at a redshift of $z=0.35$.
	The dotted lines are the theoretical predictions in linear theory,
	and the solid lines show the mean power spectra with errors on the mean in the simulations.
	The power spectra are computed for dark matter but not halos,
	and the errors are evaluated in the Gaussian approximation without the detector or residual CMB noise, $R_{\rm N}=0$.
	As a fiducial parameter, we use the value of the effective optical depth 
	$\tau_{\rm T} = (1.4 \pm 0.5)\times10^{-4}$ from the Planck result~\cite{Planck2016A&A...586A.140P}.
	The deviations from linear theory predictions occur even around $k\sim 0.05\ h{\rm Mpc}^{-1}$
	in the relative pairwise kSZ power spectrum.
	}
	\label{fig:pk}
\end{figure}

\section{Results}
\label{Sec:Results}

Having shown the power spectrum covariance of relative pairwise kSZ temperatures with the noise contributions (Eq.~(\ref{kSZCov})),
we now turn to our main goal, to understand how the relative pairwise kSZ power spectrum provides cosmological information.
To do this, we compute a cumulative signal-to-noise ratio (cumulative information)
and conduct a Fisher matrix analysis~\cite{Tegmark1997ApJ...480...22T,Font-Ribera2014JCAP...05..023F}
to see how cosmological information is lost from the relative pairwise kSZ power spectrum on small scales due to the noise
and how future kSZ measurements improve constraints on cosmological parameters,
especially for the growth rate of structure $f$ and the expansion rate of the Universe $H$.

\subsection{Relative pairwise kSZ power spectrum}

The first step to making parameter estimates from the simulations is
to determine the galaxy and relative pairwise kSZ power spectra for each realization.
In linear theory, using Eqs.~(\ref{P_n2}) and (\ref{PkSZP}),
we derive the pairwise kSZ power spectrum 
\begin{eqnarray}
	  P_{\rm kSZ}(k,\mu) &=& \left( \frac{T_0\tau_{\rm T}}{c} \right)
	  \left( i\frac{aHf}{\hat{k}\cdot\hat{n}} \right) \frac{\partial}{\partial f} 
	  \left( b + f \mu^2 \right)^2 D^2 P_{\rm lin}(k) \nonumber \\
	  &=& \left( \frac{T_0\tau_{\rm T}}{c}  \right)2iaHf\mu\left( b + f\mu^2 \right) D^2 \frac{P_{\rm lin}(k)}{k},
\end{eqnarray}
where $\left( b + f\mu^2\right)$ is the linear Kaiser factor~\cite{Kaiser1987MNRAS.227....1K},
$\mu = \hat{k}\cdot\hat{n}$, $b$ is the linear bias, $P_{\rm lin}$ is the linear matter power spectrum,
the linear growth function $D$ is normalized to unity at present, $f = d \ln D/d \ln a$ is the linear logarithmic growth rate.
Then, the pairwise kSZ power multipoles are given by
\begin{eqnarray}
	  P_{ {\rm kSZ}, \ell=1}(k) &=& 2iaH\left(  \frac{T_0\tau_{\rm T}}{c} \right) \left( f b + \frac{3}{5}f^2 \right) \frac{P_{\rm lin}(k)}{k} \nonumber \\
	  P_{ {\rm kSZ}, \ell=3}(k) &=& 2iaH\left(  \frac{T_0\tau_{\rm T}}{c} \right) \left( \frac{2}{5}f^2 \right) \frac{P_{\rm lin}(k)}{k}.
\end{eqnarray}
It is common that the linear matter power spectrum is normalized by 
$\sigma_8$, which is the current rms linear matter fluctuation on scales of $8\, h^{-1}\, {\rm Mpc}$,
and that the linear bias $b$ and the linear growth rate $f$ are rewritten as $b\sigma_8$ and $f\sigma_8$.
Thus, the pairwise kSZ power spectrum (correlation function) dipole and octopole are proportional to
$\tau_{\rm T}\left( (f\sigma_8)(b\sigma_8) + \frac{3}{5}(f\sigma_8)^2 \right)$ and $\tau_{\rm T}(f\sigma_8)^2$, respectively.
For the kSZ-only analysis, there is strong degeneracy among $b\sigma_8$, $f\sigma_8$, and $\tau_{\rm T}$.
We can define two parameters $\sqrt{\tau_{\rm T}}b\sigma_8$ and $\sqrt{\tau_{\rm T}}f\sigma_8$ and could constrain them from the kSZ-only analysis,
but it would be difficult to extract cosmological information from them; therefore, we need the joint analysis of the galaxy and kSZ power spectra.
In previous works (e.g.,~\cite{Keisler2013}), it is pointed out that the kSZ signal (dipole) is sensitive to $f\sigma_8^2$
and could be used to break the degeneracy between $f$ and $\sigma_8$ in the RSD analysis if we can know the bias parameter $b$ from galaxy clustering.
However, since the RSD analysis probes $b\sigma_8$ and $f\sigma_8$, we can not break the degeneracy between $f$ and $\sigma_8$ from the joint analysis of the galaxy and kSZ power spectrum. In what follows, we abbreviate $b\sigma_8$ and $f\sigma_8$ to $b$ and $f$ for brevity.

\begin{figure}[t]
	\centering
	\includegraphics[width=\columnwidth]{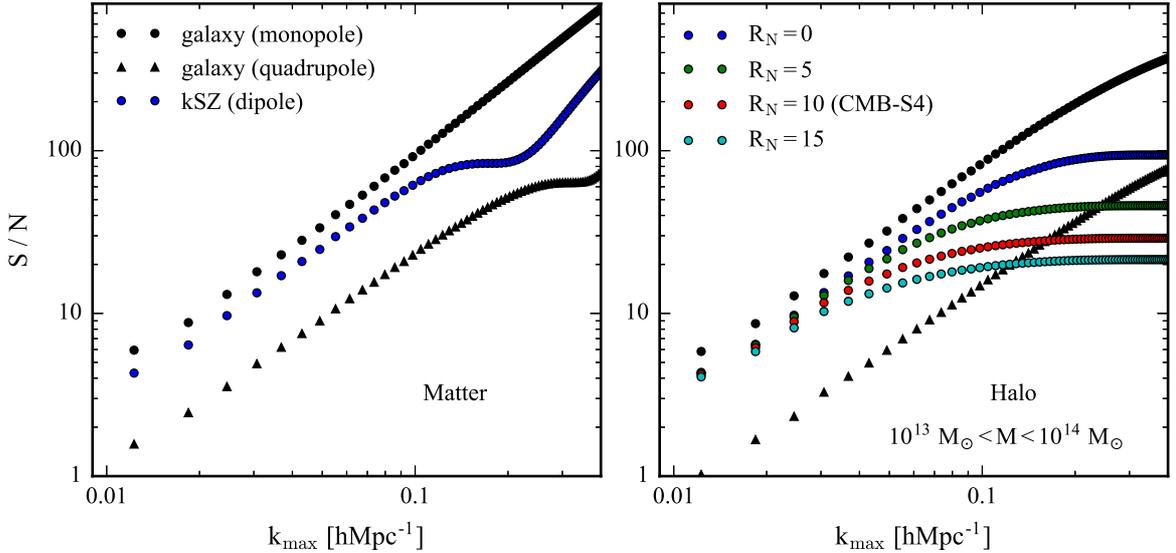}
	\caption{Cumulative signal-to-noise ratios as a function of $k_{\rm max}$ in redshift space
	for dark matter particles (left) and halos (right),
	where the power spectrum information over $2\pi/L \leq k \leq k_{\rm max}$ is included ($L$ is the box size).
	Note that the S/N amplitudes are for the simulation box volume $V = 1\ h^{-3}{\rm Gpc}^3$.
	The colored symbols show the S/N for the dipole term of the relative pairwise kSZ power spectrum,
	while the black symbols in each panel show the S/N for the monopole and quadrupole of the galaxy power spectrum,
	which approximately scale as ${\rm S/N}\propto k_{\rm max}^{3/2}$ due to ignoring non-Gaussian parts of the covariance matrices.
	The noise level of CMB-S4 roughly corresponds to $R_{\rm N}=10$.
	The inverse of S/N of the kSZ dipole power estimates the uncertainty of 
	the optical depth $\Delta \tau_{\rm T}/\tau_{\rm T}$ after fixing all the other parameters.
	The computed $\Delta \tau_{\rm T}/\tau_{\rm T}$ in our noise model (Eq.~(\ref{PkSZP})) 
	is consistent with the Planck result~\cite{Planck2016A&A...586A.140P} 
	that presents the $2.2 \sigma$ detection using the pairwise momentum estimator for $8$ arcmin aperture (see text for further details).
	}
	\label{fig:sn1}
\end{figure}

Figure~\ref{fig:pk} shows the multi-pole moment of the mean power spectra from the theoretical predictions in linear theory (dotted lines)
and the mean measured power spectra (solid lines) in the simulations
with error bars showing the scatter between individual realizations at a redshift of $z=0.35$.
To obtain the same signs and dimensions as the two-point correlation functions (Eq.~(\ref{Xi_legendre})),
in this figure we define and plot the following power spectra
\begin{eqnarray}
	  \Delta_{\ell}(k) &=& i^{\ell}\frac{k^3}{2\pi^2} P_{\ell}(k), \nonumber \\
	  \Delta_{\rm kSZ, \ell}(k) &=& i^{\ell}\frac{k^3}{2\pi^2} P_{\rm kSZ, \ell}(k),
\end{eqnarray}
where $\Delta_{\ell}$ is dimensionless, and $\Delta_{\rm kSZ,\ell}$ has the dimension of temperature $[\mu K]$.
The power spectra are computed for dark matter but not halos, and the errors are evaluated in the Gaussian approximation 
(Eqs.~(\ref{Gaussian_C1}), (\ref{Cov_g}), and (\ref{kSZCov})).
While the monopole ($P_{\ell=0}$) and the quadrupole ($P_{\ell=2}$) of the galaxy power spectrum are often discussed in some recent studies of galaxy clustering~\cite{Reid2012MNRAS.426.2719R,Xu2013,Anderson2014MNRAS.439...83A,Beutler2014MNRAS.443.1065B},
it has been shown in~\cite{Taruya2011PhRvD..83j3527T} that including terms up to the hexadecapole $(P_{\ell=4})$ recovers most of the information
contained in the full 2-dimensional galaxy power spectrum.
For the kSZ measurement,
we consider the dipole ($P_{\rm kSZ, \ell=1}$) and the octopole $(P_{\rm kSZ,\ell=3})$ of the relative pairwise kSZ power spectrum (Eq.~(\ref{PkSZP})).
We note that linear theory includes up to the hexadecapole ($\ell=4$) in the galaxy power spectrum
and up to the octopole ($\ell=3$) in the relative pairwise kSZ power spectrum, respectively.
Since the deviations from linear theory predictions 
are clearly present around $ k\sim0.05\ h{\rm Mpc}^{-1}$ at $z=0.35$,
at these scales linear theory will not be sufficient for an analysis of future high precision data.
A few previous studies have proposed non-linear theories for 
the relative pairwise kSZ (LOS velocity) power spectrum~\cite{Okumura2014JCAP...05..003O,Sugiyama2015arXiv150908232S}.
As discussed in the last paragraph of Sec.~\ref{Sec:Power},
the sign of $\Delta_{\rm kSZ, \ell=1}$ changes around $k\sim0.2\ h{\rm Mpc}^{-1}$ from negative to positive
due to non-linear redshift space distortions.

\subsection{Cumulative signal-to-noise ratio}
\label{Sec:SN}

As a useful way to quantify the impact of the noise in the kSZ measurement (Eq.~(\ref{kSZCov})),
we study the cumulative signal-to-noise ratio (S/N) for measuring the relative pairwise kSZ power spectrum over a range of wavenumbers.
The S/N is defined as
\begin{eqnarray}
	  \left( \frac{S}{N} \right)^2
	  &=&  \sum_{k_i,k_j < k_{\rm max}} P_{\rm kSZ, \ell}(k_i)
	  {\rm Cov}^{-1}\left( \widehat{P}_{\rm kSZ, \ell}(k_i), \widehat{P}_{\rm kSZ, \ell}(k_j) \right)P_{\rm kSZ, \ell}(k_j),
	  \label{SN}
\end{eqnarray}
where ${\rm Cov}^{-1}$ represents the inverse of the covariance matrix that is computed in the Gaussian approximation in Eq.~(\ref{kSZCov}),
and the summation is up to a given maximum wavenumber $k_{\rm max}$.
As long as the relative pairwise kSZ power spectrum do not vary rapidly within the bin widths $\Delta k = 0.006\ h{\rm Mpc}^{-1}$,
the S/N is independent of the bin widths. For comparison, we also compute the S/N for the galaxy power spectrum 
by replacing $P_{\rm kSZ,\ell}$ in Eq.~(\ref{SN}) by the galaxy power spectrum.
The inverse of the S/N roughly estimates the uncertainty of a parameter $\theta$ 
if all other parameters are fixed: $\Delta \theta \propto ({\rm S/N})^{-1}$.

\begin{figure}[t]
	\centering
	\includegraphics[width=\columnwidth]{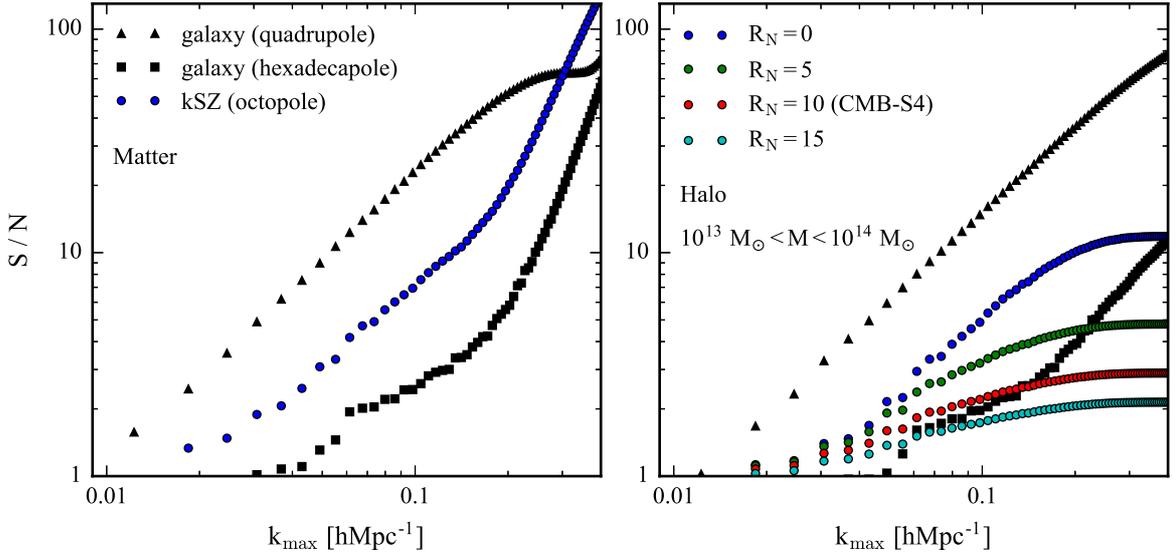}
	\caption{Same plots as Figure~\ref{fig:sn1}. 
	The black triangles are the same as those in Figure~\ref{fig:sn1}.
	These figures show the S/N for the quadrupole and hexadecapole of the galaxy power spectrum as well as
	the octopole of the relative pairwise kSZ power spectrum.
	}
	\label{fig:sn2}
\end{figure}

Figure~\ref{fig:sn1} shows the S/N for dark matter particles (left) and halos (right) as a function of $k_{\rm max}$ 
in redshift space at $z=0.35$. The S/N scales with the survey volume $V$ as ${\rm S/N} \propto \sqrt{V/(h^{-3}{\rm Gpc}^3)}$, where
the results of S/N shown here are for a volume of $V = 1\ h^{-3}{\rm Gpc}^3$.
We plot the S/N for the dipole of the relative pairwise kSZ power spectrum $P_{\rm kSZ,\ell=1}$ (colored circles),
and for comparison, also plot the S/N for the monopole (black circles) and quadrupole (black triangles) of the galaxy power spectrum,
$P_{\ell=0}$ and $P_{\ell=2}$,
which approximately scale as ${\rm S/N} \propto k_{\rm max}^{3/2}$ due to the Gaussian approximation of the covariance.
In the left panel, the S/N for $P_{\rm kSZ,\ell=1}$ and $P_{\rm \ell=2}$ become flat 
around $k_{\rm max}=0.1\ h{\rm Mpc}^{-1}$ and $k_{\rm max}=0.3\ h{\rm Mpc}^{-1}$, respectively,
because their signals become from negative to positive around these scales,
resulting in small contributions to the S/N (see also Figure~\ref{fig:pk}).
For the signal/noise levels likely for upcoming CMB experiments, the residual detector and primary CMB
noise limits the S/N for $P_{\rm kSZ}$   for modes 
in the weakly non-linear regime ($k\sim 0.1\mathchar`-0.2\ h{\rm Mpc}^{-1}$).  This is also the scale where non-Gaussian contributions to the covariance matrix begin to become important.
The flat feature appears even if $R_{\rm N}=0$ due to the shot-noise $\sigma_{\rm v}^{(2)}/\bar{n}$,
and the CMB noise contribution, which is assumed as a Gaussian random field (Sec.~\ref{Sec:Noise}), 
behaves as an enhanced shot-noise term $R_{\rm N}^2 \sigma_{\rm v}^{(2)}/\bar{n}$ as shown in Eq.~(\ref{kSZCov}).
Therefore, even in the limit of no detector noise, there is no S/N gain 
associated with using  modes at smaller scales than $k\sim 0.1\mathchar`-0.2\ h{\rm Mpc}^{-1}$.
The degradation becomes more significant with increasing the inverse signal-to-noise ratio $R_{\rm N}$:
for the result of $R_{\rm N}=10$, which roughly correspond to the forecasts for the CMB-S4 survey for 2 arcmin aperture,
the S/N is degraded by up to a factor of $2$ and $3$, respectively, 
compared to the result without the detector or residual CMB noise $R_{\rm N}=0$.
A remarkable feature in the right panel is that
until the weakly non-linear regime ($k\sim 0.1\mathchar`-0.2\ h{\rm Mpc}^{-1}$) the S/N for $P_{\rm kSZ,\ell=1}$ is larger than
that for $P_{\ell=2}$ even if $R_{\rm N}=0\mathchar`-15$.

Figure~\ref{fig:sn2} shows the S/N for the octopole of the relative pairwise kSZ power spectrum $P_{\rm kSZ,\ell=3}$ which yields from RSDs.
For comparison, we also plot the S/N for the quadrupole and hexadecapole of the galaxy power spectrum, $P_{\ell=2}$ and $P_{\rm \ell=4}$.
Similar to Figure~\ref{fig:sn1},
the noise degrades the S/N for $P_{\rm kSZ,\ell=3}$ at smaller scales than $k\sim 0.1\mathchar`-0.2\ h{\rm Mpc}^{-1}$,
and the S/N begins to become flat.
It should be noting that the S/N for $P_{\rm kSZ,\ell=3}$ is larger than that for $P_{\rm \ell=4}$ until $k\sim 0.1\mathchar`-0.2\ h{\rm Mpc}^{-1}$
for $R_{\rm N}=0\mathchar`-10$.

\begin{figure}[t]
	\centering
	\includegraphics[width=0.7\columnwidth]{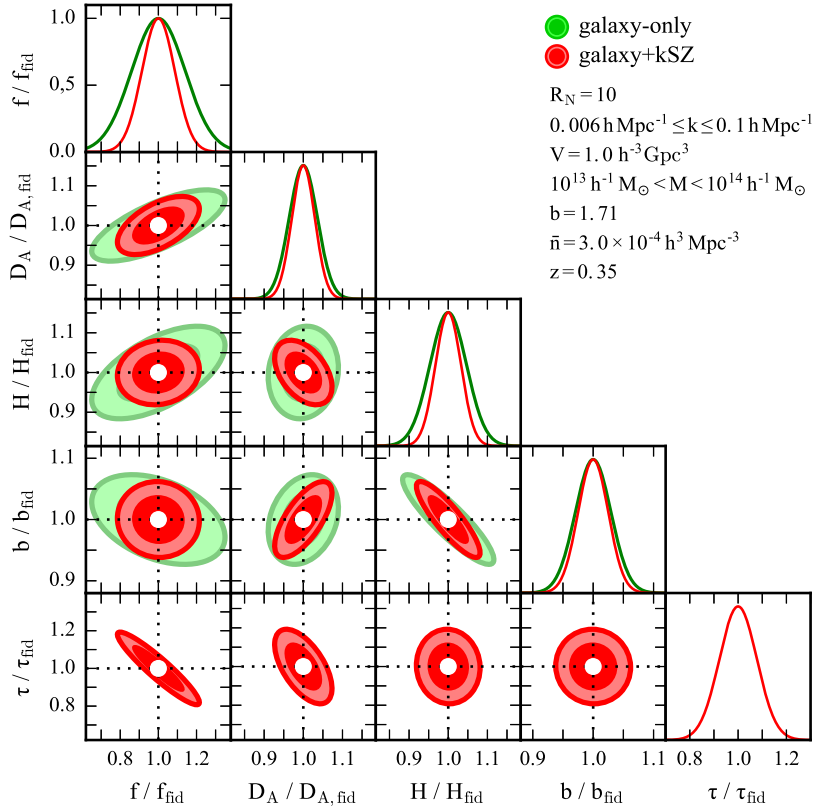}
	\caption{ 
	Confidence contours (1- and 2-$\sigma$ colored ellipses) placed on all pairs of parameters derived from our Fisher matrix analysis,
	assuming $R_{\rm N}=10$ which roughly provides the forecasts of CMB-S4 for 2 arcmin aperture.
	Constraints from the galaxy+kSZ (red regions) and galaxy-only (green regions) information are shown.
	The galaxy power spectrum does not depend on the effective optical depth $\tau_{\rm T}$ (bottom panels).
	The galaxy+kSZ information improves the galaxy-only constraints for 
	the growth rate $f(z)$, the expansion rate $H(z)$, and the effective optical depth $\tau_{\rm T}$.
	}
	\label{fig:fisher1}
\end{figure}

We note that in the covariance (Eq.~(\ref{kSZCov})) the dominant noise contribution at large scales
is $(1/V) P_{\rm s}(\kk) R_{\rm N}^2 /\bar{n}$.
This term is independent of survey volume;
therefore, the noise contribution will decrease with increasing the mean number density with fixing survey volume.

\subsection{Fisher analysis}
\label{Sec:Fisher}

\begin{figure}[t]
	\centering
	\includegraphics[width=\columnwidth]{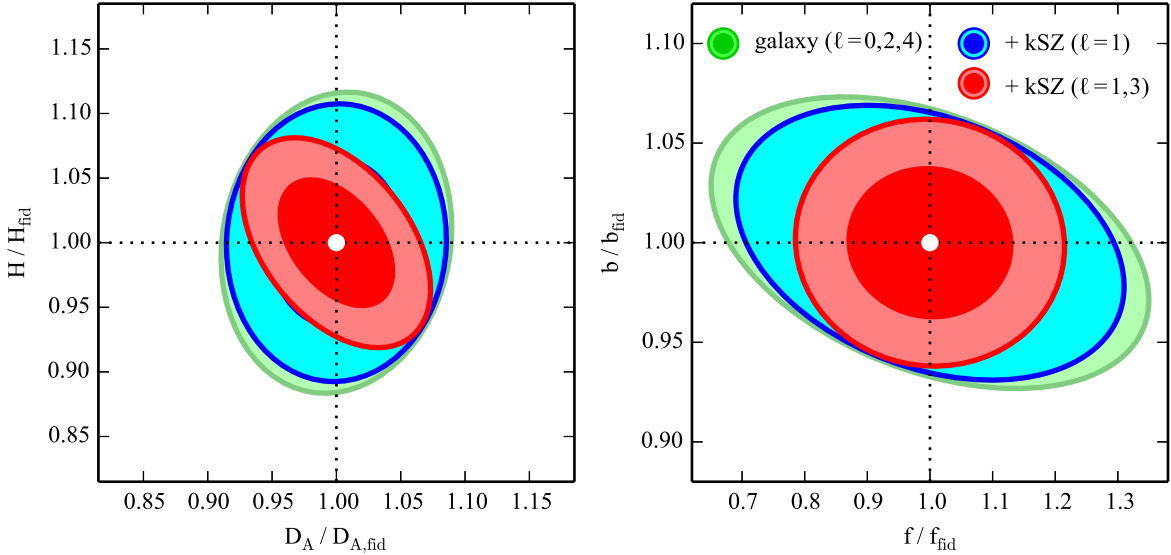}
	\caption{Same $H-D_{\rm A}$ and $b-f$ planes as those of Figure~\ref{fig:fisher1}.
	The green and red ellipses are the same as those plotted in Figure~\ref{fig:fisher1}.
	Additional confidence contours (blue regions) from the joint analysis of the galaxy-only data set
	and the dipole of the pairwise kSZ power spectrum are shown.
	Compared to the galaxy-only constraints, including both the dipole and octopole reduces the marginalized $1$-$\sigma$ errors on $H$ and $f$
	to $\sim70\%$ (from $\Delta H/H = 4.7\%$ to 3.3\%) and $\sim60\%$ (from $\Delta f/f=14\%$ to 8.6\%), respectively, for 
	the same parameters as those in Figure~\ref{fig:fisher1}.}
	\label{fig:fisher2}
\end{figure}

The Fisher matrix formalism is a standard tool for forecasting constraints on parameters of interest around a fiducial cosmology.
For the mean of a typical vector of measured quantities, $X(\theta)$,
that is predictable given parameters $\theta$, and covariance $C$, the Fisher matrix is:
\begin{eqnarray}
	  F_{ij} = \frac{\partial  X^{\rm T}}{\partial \theta_i} C^{-1}\frac{\partial  X}{\partial \theta_j},
\end{eqnarray}
where we assumed that the covariance matrix $C$ is independent of the parameters,
because the parameter dependence of $C$ becomes a sub-dominant part of the likelihood function once the parameters 
are sufficiently precisely determined~\cite{Kilbinger2006MNRAS.366..983K,Eifler2009A&A...502..721E}.
The indices $i$ and $j$ run over parameters of interest.
In the limit of Gaussian likelihood surface, 
the Cramer-Rao inequality shows that the Fisher matrix provides the minimum standard deviation on parameters,
marginalized over all the other parameters: $\Delta \theta_i \geq \left( F^{-1} \right)^{1/2}_{ii}$.
For simplicity, we assume the same fiducial cosmology as that of the simulations (see Sec.~\ref{Sec:Simulations}),
and only allow the following five parameters to vary: $\theta = \{D_{\rm A}, H, f, b, \tau_{\rm T}\}$. 
The data set consists of $X_{\rm kSZ} = \{P_{\rm \ell=0},P_{\rm \ell=2},P_{\rm \ell=4}, P_{\rm kSZ,\ell=1}, P_{\rm kSZ,\ell=3} \}$.
The $P_{\ell}$ and $P_{\rm kSZ,\ell}$ are the multi-pole moment of the galaxy and relative pairwise kSZ power spectra.
We use broadband information of the galaxy and kSZ power spectra up to some quoted $k_{\rm max}$.
Therefore, we partially use information on the Baryon Acoustic Oscillations (BAO) in galaxy clustering up to $k_{\rm max}$.
For comparison, we also conduct a Fisher matrix analysis considering only the galaxy power spectrum data set
$X_{\rm galaxy} = \{P_{\rm \ell=0},P_{\rm \ell=2},P_{\rm \ell=4} \}$ with the following four parameters $\theta = \{D_{\rm A}, H, f, b\}$.
In what follows, we refer to the data sets $X_{\rm kSZ}$ and $X_{\rm galaxy}$ as ``galaxy+kSZ'' and ``galaxy-only'', respectively.
We evaluate partial derivatives of the galaxy and kSZ power spectra
with respect to all parameters $\theta_i$ in Appendix~\ref{ap:derivative}.
For plotting of confidence ellipses in the Fisher analysis, we use the Python software ``Fisher.py''
that is available publicly~\cite{Coe2009arXiv0906.4123C}
\footnote{\url{http://www.stsci.edu/~dcoe/Fisher/}}.

\begin{figure}[t]
	\centering
	\includegraphics[width=\columnwidth]{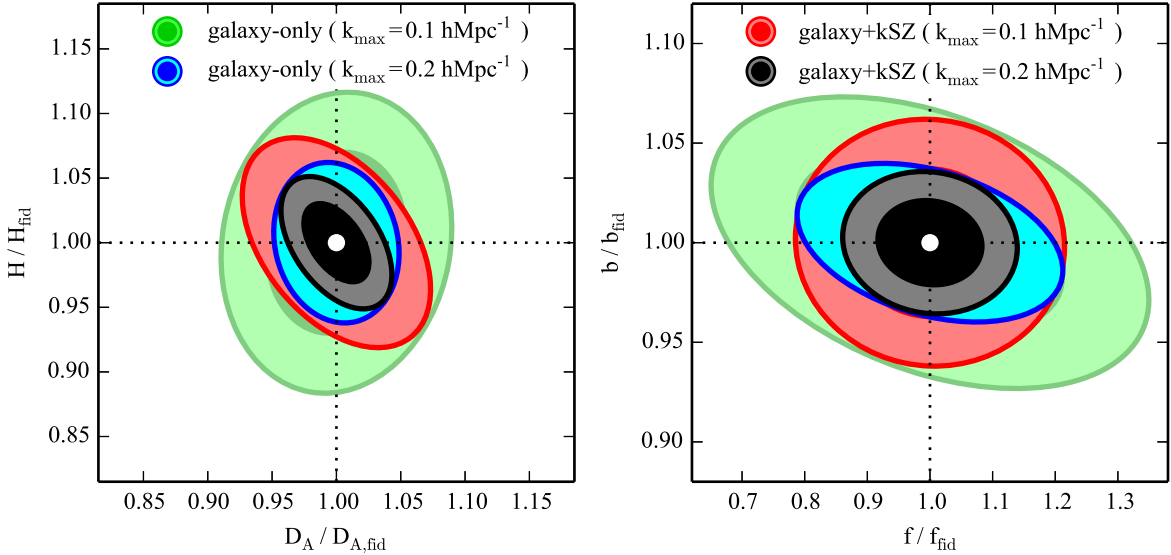}
	\caption{
	Same $H-D_{\rm A}$ and $b-f$ planes as those of Figure~\ref{fig:fisher1}.
	The green and red ellipses are the same as those plotted in Figure~\ref{fig:fisher1}.
	Additional confidence contours (blue and black regions) from the galaxy-only and galaxy+kSZ analyses 
	using broadband information up to $k_{\rm max}=0.2\ h{\rm Mpc}^{-1}$ are shown.
	Using up to $k_{\rm max}=0.2\ h{\rm Mpc}^{-1}$, the galaxy+kSZ analysis reduces the marginalized $1$-$\sigma$ errors on $H$ and $f$
	to $\sim80\%$ (from $\Delta H/H = 2.5\%$ to 2\%) and $\sim66\%$ (from $\Delta f/f = 8.5\%$ to 5.6\%), respectively.
	The galaxy+kSZ analysis using up to $k_{\rm max}=0.1\ h{\rm Mpc}^{-1}$ (red regions) improves constraints on $H$ and $f$ 
	by roughly the same amount as the galaxy-only analysis would have when using up to $k_{\rm max}=0.2\ h{\rm Mpc}^{-1}$ (blue regions).
	}
	\label{fig:fisher3}
\end{figure}

\begin{figure}[t]
	\centering
	\includegraphics[width=\columnwidth]{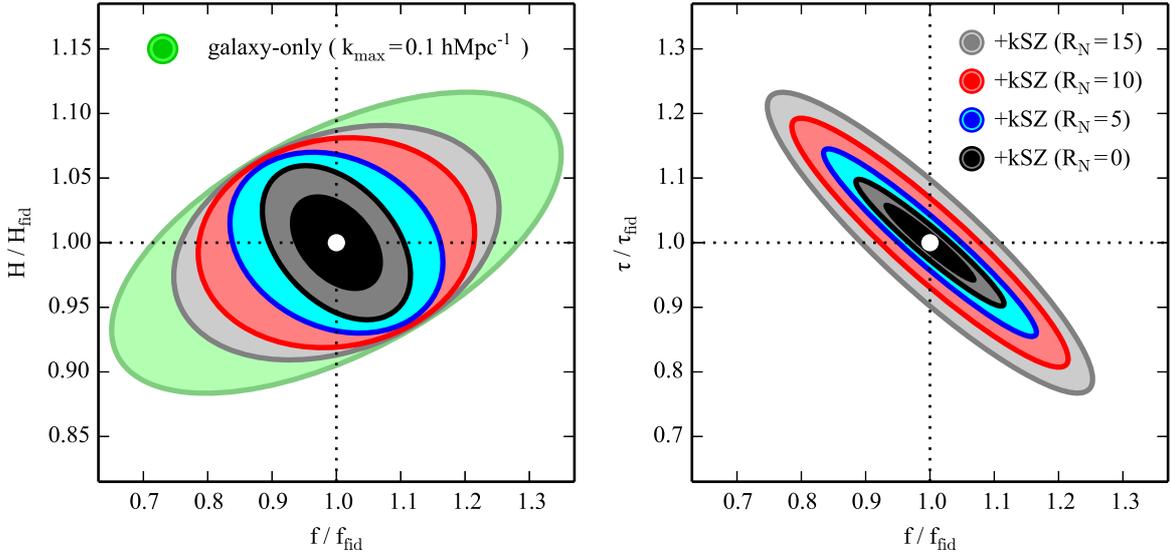}
	\caption{
	Same $H-f$ and $\tau_{\rm T}-f$ planes as those of Figure~\ref{fig:fisher1}.
	The green and red ellipses are the same as those plotted in Figure~\ref{fig:fisher1}.
	Additional confidence contours (grey, blue, and black regions) from the galaxy+kSZ analyses for $R_{\rm N}=$ 0, 5, and 15 are shown.
	$R_{\rm N}=10$ roughly provides the forecasts of the CMB-S4 survey.
	Compared to the galaxy-only analysis,
	the galaxy+kSZ analysis reduces the marginalized $1$-$\sigma$ errors on $H$ and $f$
	to $\sim50\%$, $\sim60\%$, $\sim70\%$, and $\sim77\%$ (from $\Delta H/H = 4.7\%$ to 2.4\%, 2.8\%, 3.3\%, and 3.6\%),
	and $\sim33\%$, $\sim48\%$, $\sim60\%$, and $\sim73\%$ (from $\Delta f/f = 14\%$ to 4.6\%, 6.7\%, 8.6\%, and 10.2\%),
	for $R_{\rm N}=$ 0, 5, 10, and 15, respectively.
	The effective optical depth which includes information on the missing baryon is constrained as
	3.9\%, 5.8\%, 7.6\%, and 9.4\% for $R_{\rm N}=$ 0, 5, 10, and 15.
	Note that the constraints shown here adopted a volume of $V = 1\ h^{-3}{\rm Gpc}^3$.
	Since the constraints scale as $V^{-1/2}$,
	we can obtain the rough error estimates of cosmological parameters 
	for the DESI ($V\sim10.5\ h^{-3}{\rm Gpc}^3$) and BOSS ($V\sim3.5\ h^{-3}{\rm Gpc}^3$) surveys 
	by multiplying by $1/\sqrt{10.5}\sim0.3$ and $1/\sqrt{3.5}\sim0.5$, respectively.
	}
	\label{fig:fisher4}
\end{figure}

In Figure~\ref{fig:fisher1} we show the constraints possible on all pairs of parameters in our Fisher matrix analysis,
assuming perfect knowledge of all the other parameters.
We show the parameter constraints from the galaxy+kSZ (red regions) and galaxy-only (green regions) data sets.
In this figure, it is intended to compare constraints from the galaxy-only and galaxy+kSZ measurements,
but is not meant to provide complete forecasts for future surveys.
Note that the constraints shown here are for a volume of $V = 1\ h^{-3}{\rm Gpc}^3$, where the constraints scale as $(h^{-3}{\rm Gpc}^3/V)^{1/2}$.
Since the S/N for the relative pairwise kSZ power spectrum becomes flat around $k_{\rm max} \sim 0.1\ h{\rm Mpc}^{-1}$, 
we focus on $k_{\rm max} = 0.1\ h{\rm Mpc}^{-1}$.
We assume that the inverse S/N per galaxy, $R_{\rm N}=10$.
This is consistent with forecasts for the CMB-S4 survey.
These plots can be compared to those in Figures~\ref{fig:fisher2}, \ref{fig:fisher3}, and \ref{fig:fisher4}
which show confidence contours for different parameter values in the same plots as Figure~\ref{fig:fisher1}.
Since the galaxy power spectrum does not depend on the effective optical depth $\tau_{\rm T}$, 
the galaxy-only analysis does not constrain it (green regions in the bottom panels).
The galaxy power information can break the strong degeneracy between $f$ and $\tau_{\rm T}$ in the relative pairwise kSZ power spectrum.
As expected, the galaxy+kSZ information improves the galaxy-only constraints for the growth rate $f$ and the expansion rate $H$.

To illustrate the importance of the RSD effect in the relative pairwise kSZ power spectrum,
Figure~\ref{fig:fisher2} plots the constraints from the galaxy-only (green regions) and galaxy+kSZ analyses
in the $H-D_{\rm A}$ (left) and $b-f$ (right) planes,
where kSZ information includes only the dipole (blue regions) or both the dipole and octopole (red regions).
The octopole is more sensitive to the expansion rate $H$ than the dipole in the sense that the octopole is 
a higher multi-pole moment than the dipole.
Since the octopole does not depend on the linear bias parameter $b$ (see Appendix~\ref{ap:derivative} and \cite{Sugiyama2015arXiv150908232S})
like the hexadecapole of the galaxy power spectrum,
the octopole is also sensitive to the growth rate $f$ than the dipole whose amplitude 
depends on the growth rate, linear bias, and effective optical depth.
Compared to the galaxy-only constraints (green regions),
including both the dipole and octopole (red regions) reduces the marginalized $1$-$\sigma$ errors on $H$ and $f$
to $\sim70\%$ (from 4.7\% to 3.3\%) and $\sim60\%$ (from 14\% to 8.6\%), respectively,
for the parameters used in Figure~\ref{fig:fisher1}.

Figure~\ref{fig:fisher3} shows the dependence of maximum wavenumber $k_{\rm max}$ used for the Fisher analysis.
The improvement from using broadband information depends strongly on the $k_{\rm max}$,
where we use two simple choices of $k_{\rm max}$, $0.1$ and $0.2$ $h{\rm Mpc}^{-1}$.
As well known, including small scale information improves the errors on all parameters.
The galaxy+kSZ analysis using broadband information up to $k_{\rm max}=0.1\ h{\rm Mpc}^{-1}$ (red regions)
improves constraints on $H$ and $f$ 
by roughly the same amount as the galaxy-only analysis would have when using up to $k_{\rm max}=0.2\ h{\rm Mpc}^{-1}$ (blue regions).
While we ignored non-Gaussian parts in the covariance, there are non-vanishing contributions in the off-diagonal elements in the covariance 
from the non-Gaussian errors
around $k=0.2\ h{\rm Mpc}^{-1}$~\cite{Rimes2005MNRAS.360L..82R,Rimes2006MNRAS.371.1205R,Takahashi2009,Kiessling2011MNRAS.416.1045K,
Schaan2014PhRvD..90l3523S}. 
In this case, the Fisher ellipses using up to $k_{\rm max}=0.2\ h{\rm Mpc}^{-1}$ (blue and black regions) may become far wider than shown in Figure~\ref{fig:fisher3}.

Also of interest is the potential benefit of more optimized measurements of the kSZ signal
which may reduce the inverse signal-to-noise ratio, $R_{\rm N}$.
Figure~\ref{fig:fisher4} shows the constraints from the galaxy+kSZ analysis for 
$R_{\rm N}=$ $0$, $5$, $10$, and $15$ in the $H-f$ (left) and $\tau_{\rm T}-f$ (right) planes.
Note that $R_{\rm N}=10$ roughly provides the forecast of the CMB-S4 survey 
for $\theta_{\rm F}=2$ arcmin (Sec.~\ref{Sec:Noise}).
Compared to the galaxy-only analysis, the kSZ information leads to a reduction of the marginalized $1$-$\sigma$ errors on $H$ 
by $\sim50\%$, $\sim60\%$, $\sim70\%$, and $\sim77\%$ for $R_{\rm N}=$ $0$, $5$, $10$, and $15$, respectively.
Similarly, the errors on $f$ are reduced to $\sim33\%$, $\sim48\%$, $\sim60\%$, and $\sim73\%$.
In addition to cosmologically interesting parameters,
the effective optical depth which includes information on the missing baryon~\cite{Hernandez2015} 
can be constrained: $\Delta \tau_{\rm T}/\tau_{\rm T}=$ 3.9\%, 5.8\%, 7.6\%, and 9.4\% for $R_{\rm N}=$ 0, 5, 10, and 15.

\section{Conclusions}
\label{Sec:Conclusion}

Over the past few years, cosmologists have been able to make the first detections of the kSZ effect
by combining galaxy data with measurements from ACT, Planck and SPT
\cite{Hand2012,Planck2016A&A...586A.140P,Schaan2016,Hill2016,Soergel2016}.
With the sensitivity of CMB experiments improving in the next few years,
we will soon be able to make accurate measurement of the kSZ power spectrum.
This paper emphasizes the potential scientific return from these measurements.  Unlike
previous studies, our analysis emphasizes the important information in the anisotropy of the kSZ power spectrum.

For our analysis, we have derived a simple analytic form for the power spectrum covariance of the relative pairwise kSZ temperature in redshift space
under the assumption that the noise in the filtered maps is uncorrelated between the positions of galaxies in the survey (Equation~(\ref{kSZCov})).

For the upcoming CMB-S4 survey, we have estimated 
the cumulative S/N ratios 
for measuring the relative pairwise kSZ power spectrum over $0.006\ h{\rm Mpc}^{-1} < k < 0.4\ h{\rm Mpc}^{-1}$
(Figures~\ref{fig:sn1} and \ref{fig:sn2}).
The S/N reveals two striking features:
the cumulative S/N becomes flat in weakly non-linear regime ($k\sim 0.1\mathchar`-0.2\ h{\rm Mpc}^{-1}$)
due to the expected noise level of detector noise, residual primordial CMB contamination and residual foregrounds.
For the CMB-S4 survey, 
the S/N for both the dipole and octopole of the kSZ power spectrum are degraded by a factor of $3$ relative to the S/N for a noiseless sky map.
Second remarkable feature is that the S/N for the dipole and octopole of the kSZ power spectrum
are larger than those for the quadrupole and hexadecapole of the galaxy power spectrum 
until the weakly non-linear regime ($k\sim 0.1\mathchar`-0.2\ h{\rm Mpc}^{-1}$).
Thus, on large scales we can use information on kSZ measurements in redshift surveys 
to constrain cosmological parameters complementary to galaxy clustering measurements.

Including the octopole has a significant improvement on the predicted cosmological constraints,
especially for the growth rate of structure $f$ and the expansion rate of the Universe $H$ (Figure~\ref{fig:fisher4}).
For the CMB-S4 survey,
the joint analysis of the galaxy and kSZ power spectra using broadband information up to $k_{\rm max}=0.1\ h{\rm Mpc}^{-1}$ 
reduces the marginalized $1$-$\sigma$ errors on $H$ and $f$ to $\sim 50\mathchar`-70\%$ compared to a galaxy-only analysis that uses information at  $k < 0.1 \ h{\rm Mpc}^{-1}$ .  On smaller scales, non-linearities, baryon feedback and non-Gaussian covariances become important for galaxy surveys.  

Our analysis emphasizes the value of kSZ measurements for studies of dark energy.  This strengthens the motivation for CMB-S4 making small scale measurement of CMB temperature fluctuations over regions of the sky covered by upcoming redshift surveys.  While this paper emphasized the upcoming DESI survey, we anticipate that we would reach similar conclusions for the WFIRST and Euclid spectroscopic surveys.

\acknowledgments

We thank E.Schaan and A.Kusaka for useful comments.
Numerical computations were carried out on Cray XC30 at Center for Computational Astrophysics, National Astronomical Observatory of Japan.
NSS was supported by a grant from the Japan Society for the Promotion of Science (JSPS) (No. 24-3849 and No. 28-1890).
NSS acknowledges financial support from Grant-in-Aid for Scientific Research from the JSPS Promotion of Science (25287050) 
and from a MEXT project ``Priority issue 9 to be tackled by using post-K computer''. 
T.O. was supported by Grant-in-Aid for Young Scientists (Start-up) from the Japan Society for the Promotion of Science (JSPS) (No. 26887012).
D.N.S. was partially supported by NSF  grant AST-1311756 and NASA grants NNX12AG72G and NNX14AH67G.

\appendix

\section{Conservation law}
\label{ap:conservation}

We show the relation between two-point correlation function of density fluctuations and the mean pairwise velocity
by the pair conservation equation and isotropic symmetry assumption~\cite{Peebles1976Ap&SS..45....3P,Davis1977,Peebles1980lssu.book.....P}.

In a discrete particle description, the number density is given by
\begin{eqnarray}
	  \rho(t,\xx) = \sum_{i=0}^{N-1} \delta_{\rm D}\left( \xx - \xx_i(t) \right),
\end{eqnarray}
where $\xx_i(t)$ is the three-dimensional position at object $i$.
Generally, the number of objects $N$ may be time-dependent: $N=N(t)$.
Assuming the conservation of the number of objects, $N={\rm const.}$, 
the time-derivative of the number density field leads to the continuity equation
\begin{eqnarray}
	  \frac{\partial}{\partial t} \rho(t,\xx) &=& 
	  \frac{\partial}{\partial t} \sum_{i=0}^{N-1} \int \frac{d^3k}{(2\pi)^3} e^{i\kk\cdot \left( \xx -  \xx_i(t) \right) } \nonumber \\
	  &=& 
	  \sum_{i=0}^{N-1} \int \frac{d^3k}{(2\pi)^3} \left[ -i\kk\cdot\vv_i(t) \right] e^{i\kk\cdot \left( \xx -  \xx_i(t) \right) } \nonumber \\
	  &=& 
	 -\nabla\cdot \sum_{i=0}^{N-1} \int \frac{d^3k}{(2\pi)^3} \vv_i(t) e^{i\kk\cdot \left( \xx -  \xx_i(t) \right) } \nonumber \\
	  &=& 
	  -\nabla\cdot \sum_{i=0}^{N-1} \vv_i(t) \delta_{\rm D}\left( \xx - \xx_i(t) \right) \nonumber \\
	  &=& - \nabla \cdot \left( \rho(\xx)\vv(\xx) \right),
\end{eqnarray}
where we used $\delta_{\rm D}(\xx) = \int \frac{d^3k}{(2\pi)^3} e^{i\kk\cdot\xx}$ and $\vv_i = \dot{\xx}_i$.
The continuity equation relates the number density to the density-weighted velocity (momentum) $\pp=\rho\vv$:
\begin{eqnarray}
	  \frac{\partial}{\partial t}\rho(\xx) + \nabla\cdot\pp(\xx)=0.
\end{eqnarray}
Similar to the density field,
the particle description of the three-dimensional two-point correlation function of the density field is given by
\begin{eqnarray}
	1 +   \xi(\rr) = \sum_{i,j} \delta_{\rm D}\left( \rr - \left( \xx_i - \xx_j \right) \right).
\end{eqnarray}
Then, we assume the pair conservation and obtain~\cite{Peebles1976Ap&SS..45....3P,Davis1977,Peebles1980lssu.book.....P,Fisher1994MNRAS.267..927F,Juszkiewicz1998ApJ...504L...1J,Juszkiewicz1999ApJ...518L..25J}
\begin{eqnarray}
	  \frac{\partial}{\partial t} \xi(\rr) = - \nabla \cdot \left( \left( 1 + \xi(\rr) \right) \vv_{\rm pair}(\rr)  \right)
\end{eqnarray}
where the mean relative peculiar velocity averaged over pairs at separation $r$, so-called {\it mean pairwise velocity},
$v_{\rm pair}(r)$ is defined as,
\begin{eqnarray}
	  \sum_{i,j} \left[ \vv_i - \vv_j \right]\delta_{\rm D}\left( \rr - \left( \xx_i - \xx_j \right) \right)
	  \equiv \left( 1 + \xi(\rr) \right) \vv_{\rm pair}(\rr).
\end{eqnarray}
Finally, assuming isotropic symmetry $\vv_{\rm pair} = v_{\rm pair}(\rr) \rr/r$ leads to
\begin{eqnarray}
	  \frac{\partial}{\partial t}\xi(\rr) + \frac{1}{r^2}\frac{\partial}{\partial r}\left( r^2 \left( 1 + \xi(\rr) \right)v_{\rm pair}(\rr) \right)
	  = 0,
\end{eqnarray}
and
\begin{eqnarray}
	  v_{\rm pair}(\rr) = -\frac{\int_0^r dr' r'^2 \frac{\partial}{\partial t} \xi(\rr')}{r^2 \left(  1 + \xi(\rr)  \right)}.
\end{eqnarray}

\section{Partial derivatives in a Fisher analysis}
\label{ap:derivative}

To conduct a Fisher matrix analysis,
we evaluate partial derivatives of the galaxy and kSZ power spectra
with respect to the following  parameters: $\{D_{\rm A}, H, f, b, \tau_{\rm T}\}$,
where $D_{\rm A}$, $H$, $f$, $b$, and $\tau_{\rm T}$ are the diameter distance, expansion rate, linear growth rate of structure, 
linear bias, and the effective optical depth.
For $D_{\rm A}$, $H$, and $\tau_{\rm T}$,
we can straightforwardly obtain the partial derivatives of the power spectra
with respect to these parameters through Eqs.~(\ref{Geo1}), (\ref{Geo2}), and (\ref{T_s}).
For the linear bias parameter $b$, we define the linear Kaiser factor for the $n$-th moment of the 
LOS relative pairwise velocity power spectrum $K_{\ell}^{(n)}$ as
the ratio between the linear power spectra for halos in redshift space and for matter in real space:
\begin{eqnarray}
	  K_{\ell}^{(n)} = \frac{P_{\rm lin,\ell}^{(n)}|_{\rm halo,redshift}}{P^{(n)}_{\rm lin,\ell}|_{\rm matter,real}},
\end{eqnarray}
where we obtain
\begin{eqnarray}
	  K_{\ell=1}^{(1)} &=& \left( b + \frac{3}{5}f \right)  , \nonumber\\
	  K_{\ell=3}^{(1)} &=& \frac{2}{5}f.
\end{eqnarray}
Then, we evaluate the partial derivative of the galaxy and kSZ power spectra with respect to $b$ using the following expression
\begin{eqnarray}
	  \frac{\partial P^{(n)}_{\ell}(k)}{\partial \ln b} = \frac{\partial \ln K_{\ell}^{(n)}(b)}{\partial \ln b} P^{(n)}_{\ell}(k).
\end{eqnarray}
Since we compute the galaxy and kSZ power spectra through the density and LOS velocity power spectra for halos in $N$-body simulations,
we can ignore other bias parameters to represent the scale-dependence of the power spectra for halos around a fiducial value of $b$.
For the growth rate of structure $f$,
we assume that the peculiar velocity is proportional to the growth rate $\vv \propto f$,
which is exact for $f = \Omega_{\rm m}^{0.5}$ and is a good approximation in General Relativity~\cite{Linder:2005in}.
Then, we derive the partial derivative with respect to $f$ from Eqs.~(\ref{P_nm}) and (\ref{P_n}) 
\begin{eqnarray}
	  \frac{\partial \widehat{P}_{\rm s}^{(n)(m)}(\kk)}{\partial \ln f}
	  &=& (n+m) \widehat{P}_{\rm s}^{(n)(m)}
	  + \left( i\frac{aH}{\hat{k}\cdot\hat{n}} \right)^{-1}\left( \widehat{P}_{\rm s}^{(n+1)(m)}(\kk) - \widehat{P}_{\rm s}^{(n)(m+1)}(\kk) \right), 
	  \nonumber \\
	  \frac{\partial \widehat{P}^{(n)}_{\rm s}(\kk)}{\partial \ln f}
	  &=&  n \widehat{P}^{(n)}_{\rm s}(\kk)
	  + \left( i \frac{aH}{\kk\cdot\hat{n}} \right)^{-1} \widehat{P}^{(n+1)}_{\rm s}(\kk).
	  \label{dP}
\end{eqnarray}
Note that the right-hand-sides in the above expression are measurable in simulations.
Therefore, we can directly evaluate the partial derivative of the LOS velocity power spectra with respect to $f$ using simulations.

\bibliographystyle{JHEP}
\bibliography{ms}

\end{document}